\documentclass[draft]{agujournal2019}
\usepackage{url} 
\usepackage{lineno}
\usepackage[inline]{trackchanges} 
\usepackage{soul} 
\usepackage{gensymb}
\usepackage{chemformula}
\draftfalse

\journalname{JGR: Atmospheres}

\begin{document}

%
%


\title{3D climate simulations of the Archean find that methane has a strong cooling effect at high concentrations}

%
%




\authors{J. K. Eager-Nash\affil{1},
        N. J. Mayne\affil{1},
        A. E. Nicholson\affil{1},
        J. E. Prins\affil{2,1},
        O. C. F. Young\affil{1},
        S. J. Daines\affil{3},
        D. E. Sergeev\affil{1,4},
        F. H. Lambert\affil{4},
        J. Manners\affil{5,1},
        I. A. Boutle\affil{5,1},
        E. T. Wolf\affil{6,7,8},
        I. E. E. Kamp\affil{2},
        K. Kohary\affil{1},
        and 
        T. M. Lenton\affil{3}
        }

\affiliation{1}{Department of Physics and Astronomy, University of Exeter, Exeter, EX4 4QL, UK}
\affiliation{2}{Kapteyn Astronomical Institute, University of Groningen, Landleven 12, 9747 AD Groningen, The Netherlands}
\affiliation{3}{Global Systems Institute, University of Exeter, Exeter, EX4 4QE, UK}
\affiliation{4}{Department of Mathematics and Statistics, University of Exeter, Exeter, EX4 4QF, UK}
\affiliation{5}{Met Office, FitzRoy Road, Exeter, EX1 3PB, UK}
\affiliation{6}{Laboratory for Atmospheric and Space Physics, University of Colorado Boulder, Boulder, CO, USA}
\affiliation{7}{NASA NExSS Virtual Planetary Laboratory, Seattle, WA, 98195, USA}
\affiliation{8}{NASA GSFC Sellers Exoplanet Environments Collaboration}





\correspondingauthor{Jake Eager-Nash}{J.K.Eager@exeter.ac.uk}




\begin{keypoints}
\item Methane warming peaks at a p\ch{CH4}:p\ch{CO2} ratio of 0.1, with global-mean surface warming  $<$7\,K.
\item Uneven meridional methane radiative forcing and changes to atmospheric circulation determine the equator-to-pole temperature difference.
\item 3D modelling is important to fully capture methane's cooling effect for p\ch{CH4}:p\ch{CO2} ratios greater than 0.1
\end{keypoints}

%
%

%
%


\begin{abstract} 

Methane is thought to have been an important greenhouse gas during the Archean, although its potential warming has been found to be limited at high concentrations due to its high shortwave absorption. We use the Met Office Unified Model, a general circulation model, to further explore the climatic effect of different Archean methane concentrations. Surface warming peaks at a pressure ratio p\ch{CH4}:p\ch{CO2} of approximately 0.1, reaching a maximum of up to 7\,K before significant cooling above this ratio. Equator-to-pole temperature differences also tend to increase up to p\ch{CH4}~$\leq300$\,Pa, which is driven by a difference in radiative forcing at the equator and poles by methane and a reduction in the latitudinal extend of the Hadley circulation. 3D models are important to fully capture the cooling effect of methane, due to these impacts of the circulation.

\end{abstract}

\section*{Plain Language Summary}

The Archean is a period in Earth history spanning approximately 4 to 2.5 billion years ago. During this period oxygen concentrations were much lower than today, which in turn allowed methane concentrations to be much higher during the Archean. Methane would have been produced by microbes, and depending on the exact nature of the ecosystem could have led to a wide range of atmospheric concentrations. Using a three-dimensional climate model and with various methane concentrations in the atmosphere we show that the warming effect of methane first increases and then, at high concentrations, decreases. This decrease is stronger than those found in previous studies using one-dimensional (single column) atmosphere models. We find that methane can only warm the Archean by up to 7\,K.

%
%

\section{Introduction}

Methane (\ch{CH4}) is thought to have played an important role as a greenhouse gas in warming the Archean Earth \cite{haqq-misra08,catling20}. The Archean is the geological eon spanning 4.0--2.5 billion years ago (Ga). The Archean was believed to be oxygen poor, with oxygen concentrations less than $3\times10^{-6}$ of the present atmospheric level (PAL) \cite{gregory21}, which allowed reduced gases, such as methane, to accumulate in the atmosphere \cite{catling01}. Here, we look to understand the potential role of methane, and to a lesser extent carbon dioxide (\ch{CO2}), on the Archean climate using a three-dimensional general circulation model (GCM), at various carbon dioxide concentrations. The use of a GCM further allows an understanding to be gained of the effect of methane on the global mean climate and the meridional temperature structure.

The Archean could have supported high methane concentrations at various points in its history \cite<e.g Fig. 5 in>{catling20}. A primitive pre-photosynthetic biosphere would effectively turn reductant (electron donors) from mantle inputs or oxidation of the crust into reduced organic carbon and ultimately methane \cite<e.g.>{nicholson22}. As methane will be photolysed to hydrogen at high altitude, atmospheric methane levels are then determined by the balance between surface net reductant (hence methane) input, and hydrogen escape \cite{kharecha05,goldblatt06,claire06}. \citeA{kharecha05} used a coupled ecosystem-atmosphere model to suggest that the early Archean biosphere could have sustained methane concentrations between 100 to 35,000\,ppm (equivalent to 10--3,500 Pa in surface partial pressure for a $10^5$\,Pa atmosphere). Subsequent evolution of increasingly productive photosynthetic biospheres and ultimately oxygenic photosynthesis could increase methane concentrations as they increase the net reductant input via oxidation of crustal material \cite<primarily Fe,>{walker87}. This is supported by a depletion in organic carbon-13 in deep water sediments at 2.7\,Ga \cite{eigenbrode06}. After the great oxidation event, the concentration of methane as a minority gas in an oxic atmosphere will be controlled by the balance between primarily biological production and photochemistry, where the net biological flux to the atmosphere will be controlled by ecosystem structure and biogeochemical cycling hence methane oxidation within the surface environment \cite{daines16}.

Updates to the high-resolution transmission molecular absorption database (HITRAN) have led to work finding that earlier studies have overestimated methane's potential to warm the Archean \cite{byrne15}. Previously, \citeA{haqq-misra08} found that methane, ethane and the subsequent formation of longer chain hydrocarbons can provide significant warming. However, updates to HITRAN \cite{brown13,rothman13} show an increase in the shortwave absorption by methane that was previously underestimated, leading to a decrease in tropopause radiative forcing by methane at high concentrations \cite<see discussion in>{byrne14}. This was found to lead to the greenhouse effect of \ch{CH4} becoming offset by strong shortwave absorption by methane in the atmosphere \cite{byrne15}. Using a 1D radiative convective model (RCM) \citeA{byrne15} found that surface warming due to methane is diminished at 100\,Pa (0.1\% by volume), with shortwave cooling dominating longwave heating above this. Methane shortwave radiative forcing has also been investigated for the modern Earth \cite{etminan16,collins18,byrom22}. \citeA{byrom22} found that methane's warming of the stratosphere by absorption of shortwave radiation can enhance methane longwave radiative forcing.

During the Archean, the Sun's luminosity was 75\% to 80\% of the present day value \cite{gough81}. As discussed in \citeA{charnay20}, as the Sun fuses hydrogen into helium the mean molecular weight of the Sun's core increases. The core then contracts and warms to maintain the balance between the pressure gradient and gravitational forces. This increases the rate of fusion, which causes the Solar flux to increase with time. Under present atmospheric conditions the Earth is predicted to have been globally ice covered (known as a snowball state) prior to $\sim$2.0\,Ga \cite{kasting03}. However, there is ample evidence that the Earth was not in a snowball state \cite<for example see>{feulner12,charnay20}. This was termed the faint young Sun (FYS) paradox \cite{sagan72}, with uncertainty over what kept the early Earth warm. The resolution of the FYS paradox is likely to be due to increased greenhouse gas concentrations during the Archean, particularly carbon dioxide \cite{charnay20} as well as methane.

Geological evidence has been used to predict \ch{CO2} concentrations through the Archean. Mass balance calculations from \ch{CO2} dissolved in rainwater give estimates of \ch{CO2} 10--50\,PAL (340--1,700\,Pa) at 2.7\,Ga \cite{sheldon06,driese11}, although this may be a lower limit \cite{catling20}. Another method using chemical compositions of paleosols predicts \ch{CO2} 85-510\,PAL (3,000--15,000\,Pa) at 2.77\,Ga and 78-2500\,PAL (2,000--75,000\,bar) at 2.75\,Ga \cite{kanzaki15}. A more recent approach using oxidation of fossilised micrometeorites from 2.7\,Ga \cite{tomkins16} suggests a lower limit for atmospheric \ch{CO2} of 32$\%$ (32,000\,Pa) \cite{huang21}, although debate remains over these estimates, with a lower surface pressure offering an alternative solution to explain the micrometeorite oxidation \cite{rimmer19}.

General Circulation Models (GCMs), which are three-dimensional models attempting to capture the main processes determining the planetary climate, have played an important role in understanding the climate of the Archean \cite{charnay13,wolf13,kunze14,lehir14,charnay20}. GCMs have been used to show that compared to 1D models, lower amounts of \ch{CO2} are required to maintain global surface temperatures of 15$\degree$C, and more importantly, avoid a full glaciation \cite{charnay13,wolf13}. Furthermore, the use of GCMs has found that potential reductions in land fraction and albedo during the Archean, as well as a reduction in cloud condensation nuclei may have helped to keep the early Earth warm \cite{wolf14,goldblatt20}.
GCMs have also been used to explore the potential for glaciations at the end of the Archean \cite{teitler14}, alongside being combined with models of carbon cycling to investigate the plausibility of hot Archean climates \cite{charnay17}.

In this work we extend the 1D work of \citeA{byrne15} by exploring the 3D effects of changing the methane concentration using the Met Office Unified Model ({\sc UM}) GCM. First we outline the model configurations used for our Archean-like Earth simulations using the {\sc UM} in Section~\ref{sec:methods}. The results are presented in Section~\ref{sec:results} where we demonstrate that methane has a maximum potential global warming of up to 7\,K. We then go on to show that methane changes the equator-to-pole temperature differences by changing the meridional circulation and the radiative forcing at the poles. In Section~\ref{sec:discussion} we discuss the importance of 3D modelling in understanding the meridional circulation, and the possible impact haze may have on the results. Finally, we draw conclusions in Section~\ref{sec:conclusions} and highlight future avenues for improving on this study.

\section{Modelling framework}
\label{sec:methods}

\begin{figure}
\centering
\noindent\includegraphics[width=0.6\textwidth]{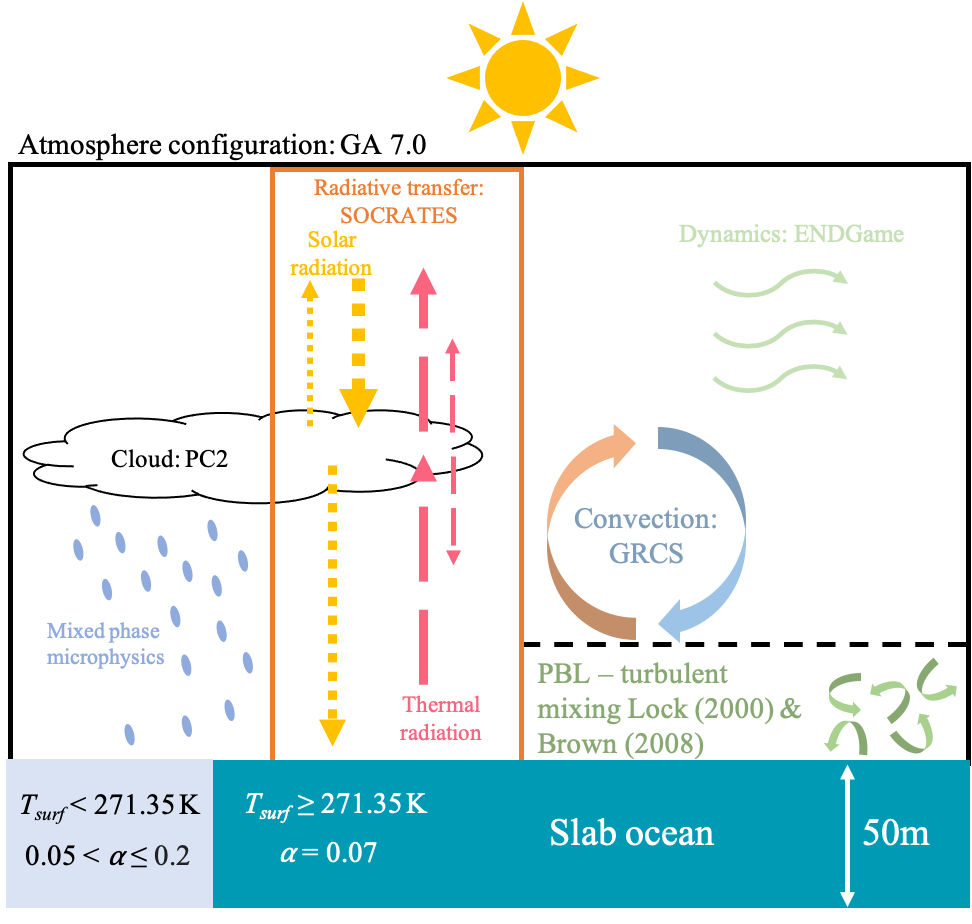}

\caption{Schematic of the different components of the the model. The GA 7.0 atmospheric configuration consists of the ENDGame dynamical core \cite{wood14}, the PC2 scheme to treat clouds \cite{wilson99,wilson08}, microphysics is based on \citeA{wilson99}, SOCRATES to treat radiative transfer, the Gregory-Rowntree Convection Scheme (GRCS) \cite{gregory90}, with turbulent mixing in the planetary boundary layer (PBL) based on \citeA{lock00} and \citeA{brown08}. The ocean is treated as a single layer slab of 50m depth with no horizontal heat transport. Sea ice albedo is treated using Equation~\ref{equ:ice_albedo}.
\label{fig:model_schematic}
}
\end{figure}

The {\sc UM} has been used extensively to study the modern Earth \cite<e.g.>{walters19,sellar19,andrews20,maher22}, and has been adapted to simulate a range of idealised Earth-like planets \cite<e.g.>{mayne14b,boutle17,lewis18,yates20,boutle20,sergeev20,eager20,sergeev22}. Here, we apply this model to the Archean, deep in the Earth's past. We use the Global Atmosphere (GA) 7.0 configuration \cite{walters19}. Dynamics are calculated using the ENDGame dynamical core \cite{wood14}, while convection is treated using a mass-flux approach based on \citeA{gregory90}. Water clouds are treated using the prognostic cloud fraction and prognostic condensate scheme (PC2) \cite{wilson08}, which incorporates mixed phase microphysics based on \citeA{wilson99}. Turbulent mixing is based on \citeA{lock00} and \citeA{brown08}. These schemes are combined and shown as a schematic in \figurename~\ref{fig:model_schematic}. Simulations have a horizontal resolution of 2.5$\degree$ in longitude by 2$\degree$ in latitude, with 38 vertical levels between the surface ($z=0$\,km) and the top-of-atmosphere ($z=40$\,km) the same as that of \citeA{eager20}. The vertical levels are quadratically stretched to enhance the resolution at the surface.

The Suite of Community Radiative Transfer codes based on Edwards and Slingo (SOCRATES) scheme treats the radiative transfer in the {\sc UM}, employing the correlated-k method (included in schematic in \figurename~\ref{fig:model_schematic}). Thermal radiation is treated via 17 bands (between 3.3\,$\mu$m-10\,mm), while solar radiation is treated by 43 bands (0.20-20\,$\mu$m) using the \texttt{sp\_lw\_17\_etw\_arcc10bar} and \texttt{sp\_sw\_43\_etw\_arcc10bar\_sun\_2.9gya} spectral files respectively. These are suitable for atmospheres dominated by a mixture of \ch{N2} and \ch{CO2} (from 1\% to 20\%), with up to 3.5\% \ch{CH4} (tested in \ref{sec:spec_file_test}), supporting surface pressures up to 10$^6$\,Pa. These include \ch{CO2} sub-Lorentzian line wings and \ch{CO2} self-broadening. Collision induced absorption is included for: \ch{N2}--\ch{CH4}, \ch{N2}--\ch{N2} and \ch{CO2}--\ch{CO2} from HITRAN \cite{karman19}, and \ch{CH4}--\ch{CO2} from \citeA{turbet20}. Line data are from HITRAN 2012 \cite{rothman13}, the same as used in \citeA{byrne15}. The solar spectrum is taken for a 2.9\,Ga Sun spectrum from \citeA{claire12}. Testing of our radiative transfer against higher resolutions for the gas mixtures used here can be found in \ref{sec:spec_file_test} and \figurename~\ref{fig:spec_file_test}, focusing on the \ch{CH4} tropopause radiative forcing.

Methane radiative forcing is calculated at the tropopause. It is the net downward total ``all-sky" radiative flux subtracted by the all-sky flux with methane switched off radiatively, which is calculated for the present model configuration without altering the climate state. Similar to the definition from the \citeA{wmo57}, the tropopause is defined as the region above 500\,hPa where the lapse rate is less than or equal to 2\,K/km for at least two consecutive vertical model levels.

The simulations were configured as an aquaplanet, using a single layer slab homogeneous flat surface as the inner boundary (planet's surface), which is based on \citeA{frierson06}. It represents an ocean surface with a 50\,m mixed layer with a heat capacity of 2.08$\times10^8$\,J/K/m$^2$, with no horizontal heat transport. The emissivity of the surface is fixed at 0.985 \cite{snyder98} and the liquid water surface albedo ($\alpha_{sea}$) is fixed at 0.07 \cite{jin04}. The assumption of an aquaplanet was made with predictions that the ocean water content may have been larger than today \cite{dong21}, and continental coverage was likely lower \cite{cawood13}. Although excluding horizontal heat transport in the ocean is inherently inaccurate, the uncertainty in continental coverage and location means that the inclusion of a dynamic ocean will also lead to further inaccuracies that may obfuscate the effects we are exploring in this work.

Sea ice albedo effect is represented by a change in albedo at 271.35\,K ($T_{threshold}$) using the HIRHAM parametrization from \citeA{liu07}, which calculates a temperature dependent sea ice albedo ($\alpha_{ice}$) as
\begin{equation}
    \boldsymbol{\alpha_{ice}=\alpha_{max}-\exp{\bigg(-\frac{T_{threshold}-T_{surf}}{2}}\bigg) \times (\alpha_{max}-\alpha_{sea}),}
    \label{equ:ice_albedo}
\end{equation}
where $T_{surf}$ is the surface temperature and $\alpha_{max}$ is the maximum sea ice albedo. A spectrally dependent $\alpha_{max}$ is used following \citeA{joshi12}, which is independent of the spectral type of the host star. For spectral bands below 0.5$\mu$m $\alpha_{max}=0.8$, while bands above 0.5$\mu$m $\alpha_{max}=0.05$. If a bands range contains 0.5$\mu$m the albedo of the band is a linear weighting of the two components. When convolved with the solar flux, in practice, our simplified scheme leads to a maximum ice albedo of approximately 0.2. Although this value is somewhat lower than expected, the simulations reproduced reasonable ice coverage for the modern Earth (not shown here). In reality, the scheme setup is compensating for missing heat transport via the ocean and the lack of a more sophisticated thermodynamic ice scheme, alongside sea ice transport. Simple non-thermodynamic ice schemes like the ones used here have been shown to result in larger climate fluctuations and generally less ice coverage than their thermodynamic counterparts \cite{poulsen04}. Therefore, in our simulations we expect our approach to result in a smaller ice albedo feedback impact, subsequently making our simulations more resistant to entering a snowball state. In the future we will continue to develop the complexity and completeness of our model, but this often comes at a price of increased difficulty in understanding specific processes. Therefore, for this work, focused on the radiative impact of methane, we elected to use the simplified scheme described.

We focus our study on 2.7\,Ga as at this time methane concentrations could have been high, as the evolution of oxygenic photosynthesis may have supported a widespread productive biosphere \cite<indicated by depletion of organic carbon-13 in marine sediments,>{eigenbrode06,daines16}. This increase in biotic oxygen production is prevented from oxidising the atmosphere by burial of oxidants in the highly insoluble form of iron oxides whilst reductant was added to the atmosphere as methane through recycling of organic carbon \cite{walker87}. Hence, there is a net oxidation of the Earth's surface and net reductant input into the atmosphere. Planetary parameters used in our simulations are presented in \tablename~\ref{tab:parameters}. The fainter Sun at 2.7\,Ga meant that the solar constant was less than the present day. From \citeA{gough81} (their Equation 1), we can estimate this to be 81$\%$ of the modern solar constant, $S_0$, of 1361\,W/m$^2$ \cite{kopp11}.

At 2.7\,Ga the Earth's rotation rate was believed to be faster than the present Earth. \citeA{williams00} predict that the rotational period at 2.45\,Ga could have been between 16.0 and 19.4\,hrs, while \citeA{bartlett16} (from their Figure 5) suggest a range of between approximately 16.0 to 19.5\,hrs at 2.7\,Ga depending on the magnitude of the lunar torque at the time. In this study we use the value of 17\,hrs, although the exact value over this range has a minimal impact on the overall results presented here. For simplicity, the modern obliquity (see \tablename~\ref{tab:parameters}) is used as it is unclear how obliquity could be significantly different to the present day, however, following the choices made by \citeA{charnay13,wolf14}, we set the eccentricity to zero.

In these simulations, we vary the methane partial pressures from 1 to 3,500\,Pa to cover the range predicted by \citeA{kharecha05}, as well as extending the lower limit in line with \citeA{byrne15}, to include potential methane concentrations from only abiotic sources, p\ch{CH4}$\approx$1\,Pa \cite{kasting05}. We also include a baseline case, without methane, as a comparison for our simulations with varying atmospheric methane concentrations. We vary the initialised \ch{CO2} surface partial pressure from 100 to 3,000\,Pa covering some of the large variation in paleosol constraints (340--75,000 Pa) and use the lower value of 100 Pa to match \citeA{byrne15}.  We vary p\ch{CO2} to cover a range of predicted \ch{CO2} abundances at the time, from 100--3,000\,Pa. For p\ch{CO2}=10,000\,Pa our simulations were significantly warmer than the modern day Earth and the UM becomes unstable for the higher methane concentrations. Therefore, we omitted these cases from our study and used p\ch{CO2}$\leq$3,000\,Pa as the upper limit.

These surface partial pressures are the values initialised in the model, while the total surface pressure, and thus individual partial pressures, can evolve from the initialised value. However, the equivalent volume mixing ratios have no spatial or temporal variation. The reason for the choice in using partial pressures is to ensure that the mass of \ch{N2} and \ch{CO2} is kept constant with variations in \ch{CH4} abundance. This is to replicate methane production by the biosphere, adding methane to the atmosphere. Surface pressures range from 1-1.035$\times10^5$\,Pa, which aligns with the upper limit of surface pressures from \citeA{som12}, as well as using similar conditions to \citeA{byrne15}.

Fixed gas mixing profiles are used in these simulations, with no chemistry included. This assumption is validated by 1D photochemical models, which show that the volume mixing ratio remains relatively constant up to 40\,km (top of model in our simulations) for carbon dioxide \cite<e.g.>{huang21} and methane \cite<e.g.>{kharecha05,gregory21} at low oxygen concentrations.

\begin{table}
\caption{The planetary and orbital parameters used for all planetary configurations, based on a 2.7\,Ga Earth. Stellar irradiance is calculated using \citeA{gough81}, based on a modern day solar constant $S_0=1361$\,W/m$^2$ \cite{kopp11}. Rotation rate is based on estimates from \citeA{williams00} and \citeA{bartlett16}. Partial pressures represent the values initialised in the model at the surface.
\label{tab:parameters}
}
\centering
\begin{tabular}{l c}
\hline
 \textbf{Parameter} & \textbf{2.7\,Ga Archean Earth} \\
\hline
Stellar irradiance (W/m$^{2}$) & 1100.8 (0.809$S_0$) \\
Day length (hours) & 17.0 \\
Eccentricity & 0  \\
Obliquity ($\degree$) & 23.44 \\
p\ch{CO2} (Pa) & 100, 300, 1000, 3,000 \\
p\ch{CH4} (Pa) & 0, 1, 3, 10, 30, 100, 300, 1000, 3,500 \\
p\ch{N2}+p\ch{CO2} (Pa) & 100,000 \\
Surface pressure & p\ch{N2}+p\ch{CO2}+p\ch{CH4} \\
\hline
\end{tabular}
\end{table}

Initial simulations with no \ch{CH4} ran for 60\,years to allow for atmospheric equilibrium to be reached. Subsequent runs with \ch{CH4} present were initialised from these simulations  and ran for another 50\,years to reach a new steady state. All the simulations were in steady state by the final 15 years, so we use this period for our analysis. Steady state was deemed to have been reached when the top of atmosphere radiative flux was in balance and the mean global surface temperature was near constant. In this work, we refer to equatorial and polar regions. The equatorial region is considered as spanning latitudes of 10\degree\,S to 10\degree\,N, while the polar regions are 70\degree\,S/N to the pole, which are averaged over these regions in figures henceforth. {\sc UM} output was processed and plotted using Python's {\sc iris} \cite{iris}, {\sc aelous} \cite{aeolus} and Matplotlib \cite{matplotlib} packages.
    
\section{Results} 
\label{sec:results}

In this section we demonstrate that methane has a maximum potential warming of up to 7\,K globally in our Archean-like simulations. Furthermore, we find that the equator-to-pole temperature difference generally increases with p\ch{CH4}, which is driven by the difference in methane radiative forcing at the equator and poles and also the decrease in circulation strength driven by upper troposphere heating by methane shortwave absorption. 

\begin{figure}
\centering
\noindent\includegraphics[width=\textwidth]{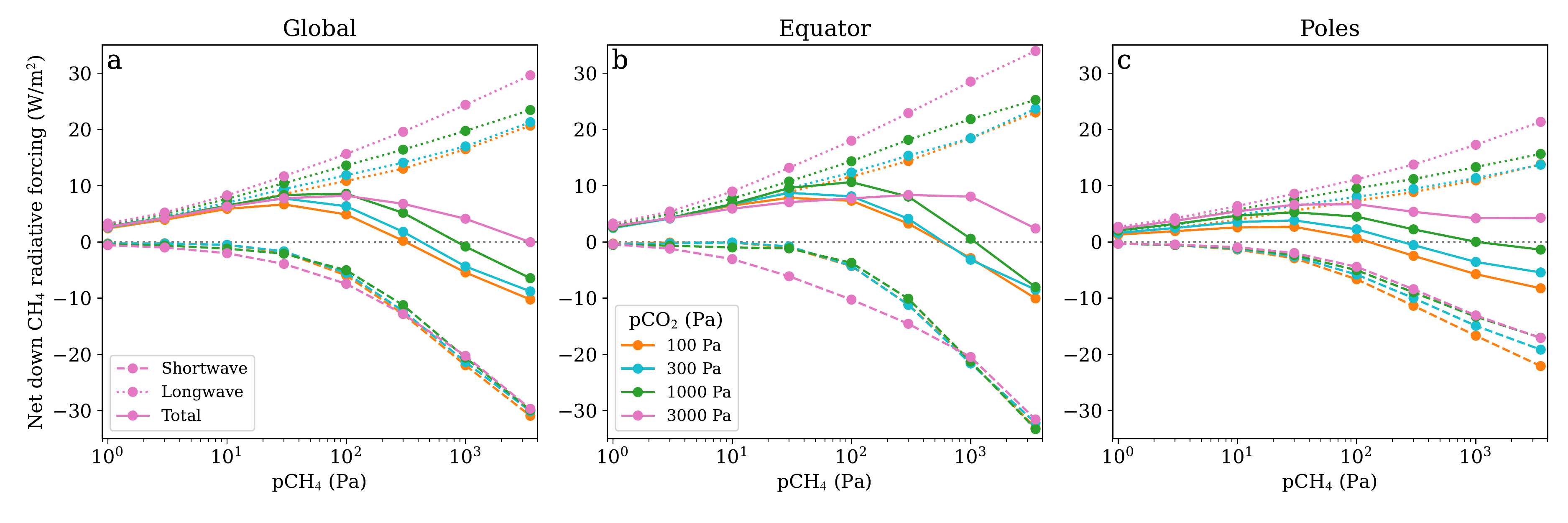}

\caption{(a) shows the global average net down radiative forcing at the tropopause for methane for shortwave (dashed), longwave (dotted) and their sum (solid). (b) and (c) show the same as (a) averaged over 10$\degree$S to 10$\degree$N as the equator in (b), and poleward of 70$\degree$N/S for the polar regions in (c).
\label{fig:CH4_rad_forcings}
}
\end{figure}

Methane radiative forcing plays an important role in driving the climate. This is shown for all of our simulations in \figurename~\ref{fig:CH4_rad_forcings}. \figurename~\ref{fig:CH4_rad_forcings}a shows the global tropopause radiative forcing by methane. As p\ch{CH4} is increased, initially total radiative forcing increases due to an increase in longwave radiative forcing (dotted lines in \figurename~\ref{fig:CH4_rad_forcings}a) causing surface temperature to increase. As p\ch{CH4} continues to increase, shortwave radiative forcing becomes more negative (dashed lines in \figurename~\ref{fig:CH4_rad_forcings}a) caused by absorption of shortwave radiation, which becomes comparable to the magnitude of longwave radiative forcing and causes methane's total radiative forcing to decrease again, driving a cooling of the surface. Similar to the results found in \citeA{byrne14}, the shortwave absorption becomes significant for p\ch{CH4}$>$10\,Pa, with the total (longwave plus shortwave) methane radiative forcing similarly having a maximum of approximately 8.5\,W/m$^2$, compared to 9\,W/m$^2$ in \citeA{byrne14}.

\figurename~\ref{fig:CH4_rad_forcings}b,c show the methane radiative forcing for equatorial and polar regions respectively. Equatorial forcing exhibits similar trends to the global average, while polar radiative forcing is of a lower magnitude, with total radiative forcing peaking at smaller p\ch{CH4} compared to the equatorial and global regions. Shortwave forcing is weaker at the poles for p\ch{CO2}$=$3000\,Pa compared to the other p\ch{CO2} values, while longwave forcing is stronger, leading to methane radiaitve forcing at the poles plateauing at p\ch{CH4}$\geq$1000\,Pa.

The effects of this change in radiative forcing by methane are now explored for these different cases by examining the difference in climate between these configurations.

\subsection{Meridional air temperature variation and the Hadley circulation}
\label{sec:maps}
\begin{figure}
\centering
\vspace*{-2cm}
\makebox[\linewidth]{
\noindent\includegraphics[width=1.2\textwidth]{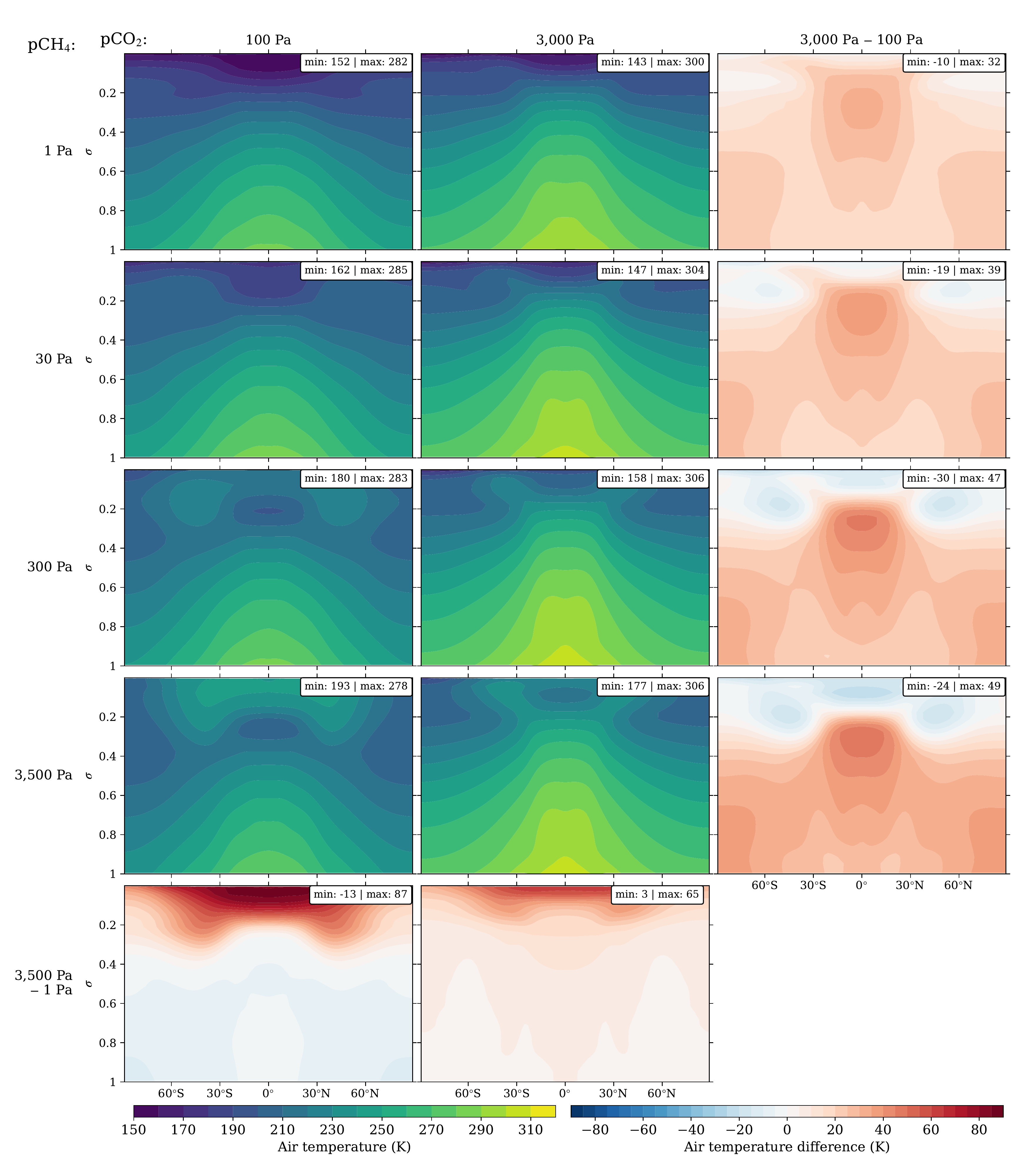}
}
\caption{Zonal averaged air temperature (colour scale), for increasing surface partial pressures of carbon dioxide from left to right, and methane from top to bottom. The right hand column shows the difference in air temperature between p\ch{CO2}$=$100\,Pa and p\ch{CO2}$=$3,000\,Pa (the left and middle columns). The bottom row shows the change in air temperature between the p\ch{CH4}$=$1\,Pa and p\ch{CH4}$=$3,500\,Pa case for a given p\ch{CO2}. In both of these cases a positive increase represents an increase in air temperature for the simulation with higher p\ch{CH4}/p\ch{CO2}. Plotted as latitude vs $\sigma$, where $\sigma$ is the pressure divided by the surface pressure. This is for a subset of the simulations, with the full grid available in \ref{sec:appendix} (\figurename~\ref{fig:T_air_maps_full_grid}). The same colour scale is used for the left and middle columns, with maximum and minimum temperatures for each simulation displayed in the top right of each sub figure. The right column and bottom row use a another colour scale that is constant between the subplots. These subplots also displays the maximum/minimum temperature differences in the top right of each sub figure.
\label{fig:T_air_maps}
}
\end{figure}

Increasing p\ch{CO2} in our Archean-like configuration leads to a familiar result in increasing temperatures globally \cite<e.g.>{charnay13}. \figurename~\ref{fig:T_air_maps} shows a subset of our simulations and reveals that increasing p\ch{CO2} from 100\,Pa to 3,000\,Pa (comparing left and middle columns) at any p\ch{CH4} leads to the zonal mean air temperatures rising globally in the troposphere, with the polar regions warming more than equatorial regions, shown by the right hand column of \figurename~\ref{fig:T_air_maps}, which shows the increase in air temperature from p\ch{CO2}$=$100\,Pa to p\ch{CO2}$=$3,000\,Pa. Minimum temperatures decrease due to the stratosphere cooling with increasing p\ch{CO2} due to more efficient cooling of the atmosphere.

\begin{figure}
\centering
\vspace*{-2cm}
\makebox[\linewidth]{
\noindent\includegraphics[width=\textwidth]{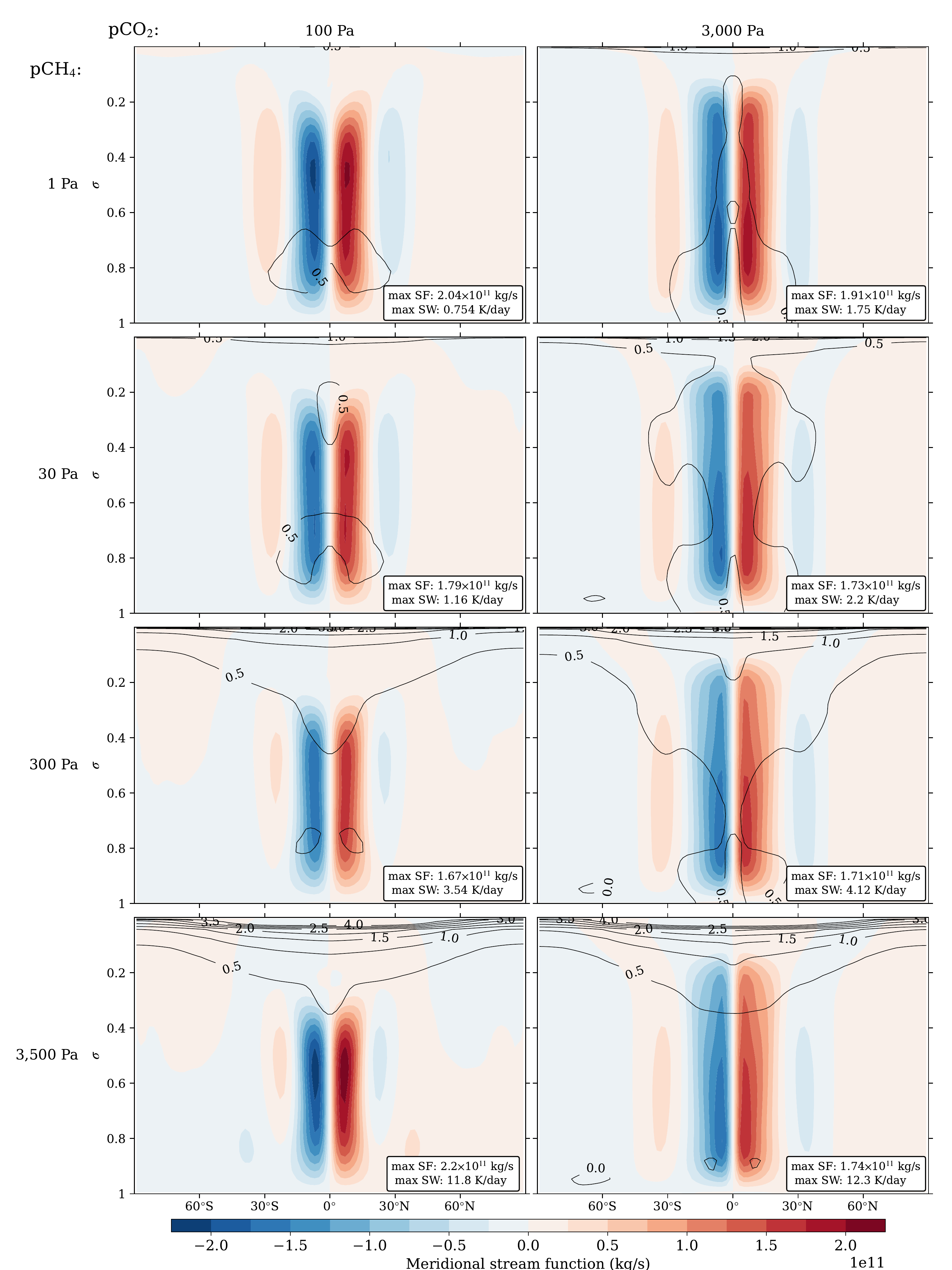}
}
\caption{Zonal averaged meridional stream functions (colour scale), for increasing surface partial pressures of carbon dioxide from left to right, and methane from top to bottom, in the same format as \figurename~\ref{fig:T_air_maps}, with the full grid available in \ref{sec:appendix} (\figurename~\ref{fig:HC_full}). Positive and negative values represent clockwise and anticlockwise circulation respectively. Contours show the heating of the atmosphere due to shortwave radiation in K/day. The same colour scale is used for each plot, with maximum values for the stream function (SF) and shortwave heating rate (SW) are shown for each simulation in the bottom right of each sub figure.
\label{fig:stream_function}
}
\end{figure}

Increasing surface p\ch{CH4} from 1\,Pa to 3,500\,Pa, leads to a smaller change in troposphere air temperatures compared to changing p\ch{CO2}, but large increases in stratospheric temperature. \figurename~\ref{fig:T_air_maps} shows that increasing p\ch{CH4} initially leads to troposphere warming, which peaks between p\ch{CH4} values of 30--300\,Pa and increasing p\ch{CH4} further leads to cooling of the troposphere at lower p\ch{CO2} and plateauing at higher p\ch{CO2} (see bottom row of \figurename~\ref{fig:T_air_maps}), which is driven by changes the methane radiative forcing (\figurename~\ref{fig:CH4_rad_forcings}a). Increasing p\ch{CH4} also leads to a warming of the stratosphere, and the formation of a stratospheric temperature inversion, visible at p\ch{CH4}$\geq$300\,Pa at p\ch{CO2}$=$100\,Pa and p\ch{CH4}$=$3,500\,Pa at p\ch{CO2}$=$3,000\,Pa in \figurename~\ref{fig:T_air_maps}. The magnitude of the warming can by seen in bottom row of \figurename~\ref{fig:T_air_maps}, which shows the change in air temperature between p\ch{CH4}$=$1\,Pa and p\ch{CH4}$=$3,500\,Pa with the stratosphere warmer by up to 87\,K at p\ch{CH4}$=$3,500\,Pa and p\ch{CO2}$=$100\,Pa. This warming of the stratosphere is caused by the increase in shortwave heating of the upper atmosphere caused by methane (shown by the contours in \figurename~\ref{fig:stream_function}). These changes in air temperature show similar trends to those presented in \citeA{byrne15}. Horizontally averaged pressure-temperature plots for the equator and poles are shown in \figurename~\ref{fig:Tqcloud}.

The Hadley circulation strength is shown in \figurename~\ref{fig:stream_function} in the form of meridional stream functions. The shorter rotational period of the planet used here (17 hour day compared to 24 hour) leads to the Hadley circulation having a reduced latitudinal depth, however there remains only three circulating cells. Additional cells are only expected to form at shorter rotational periods \cite{Kaspi15}.

The latitudinal and vertical extent of the Hadley circulation increases when increasing p\ch{CO2} as shown in \figurename~\ref{fig:stream_function} (see \figurename~\ref{fig:HC_full} for full grid of simulations). Increasing p\ch{CO2} (left to right) leads to a stronger and deeper Hadley circulation, due to more efficient cooling of the upper tropopause and increasing surface temperatures increasing specific humidity. This is shown by the Hadley cells highlighted by the stream function intensifying, extending in latitudinal breadth and increasing the range of pressures they cover as p\ch{CO2} is increased.

Increasing p\ch{CH4} leads to a decrease in the latitudinal and vertical extent of the Hadley circulation, due to methane's shortwave heating of the upper troposphere for p\ch{CO2}~$<$~3000\,Pa, while at p\ch{CO2}~$=$~3000\,Pa the Hadley cell extent remains similar. For p\ch{CO2}~$=$~100\,Pa, \figurename~\ref{fig:stream_function} shows that at lower values of p\ch{CH4} shortwave heating (black contour lines) predominantly takes place in the lower troposphere, while at higher p\ch{CH4} the shortwave heating is largest in the upper troposphere. This is caused by high p\ch{CH4} values increasing the shortwave absorption of the atmosphere. This results in the upper troposphere being more stable against convection at higher p\ch{CH4}, and reduces the tropopause height and subsequently the depth and strength of the Hadley circulation (also see lapse rates for a low p\ch{CO2} and high p\ch{CH4} case and a high p\ch{CO2} and low p\ch{CH4} in \figurename~\ref{fig:lapse_rate}). For p\ch{CO2}~$=$~3000\,Pa, efficient cooling of the upper troposphere by \ch{CO2} counteracts methane heating in the stratosphere and upper troposphere (compare warming due to methane in the bottom two panels of \figurename~\ref{fig:T_air_maps}), allowing the vertical extent of the Hadley circulation to remain similar for all p\ch{CH4} values.

\subsection{The global effect of methane on climate}
\label{sec:global_methane_effects}

\begin{figure}
\centering
\noindent\includegraphics[width=\textwidth]{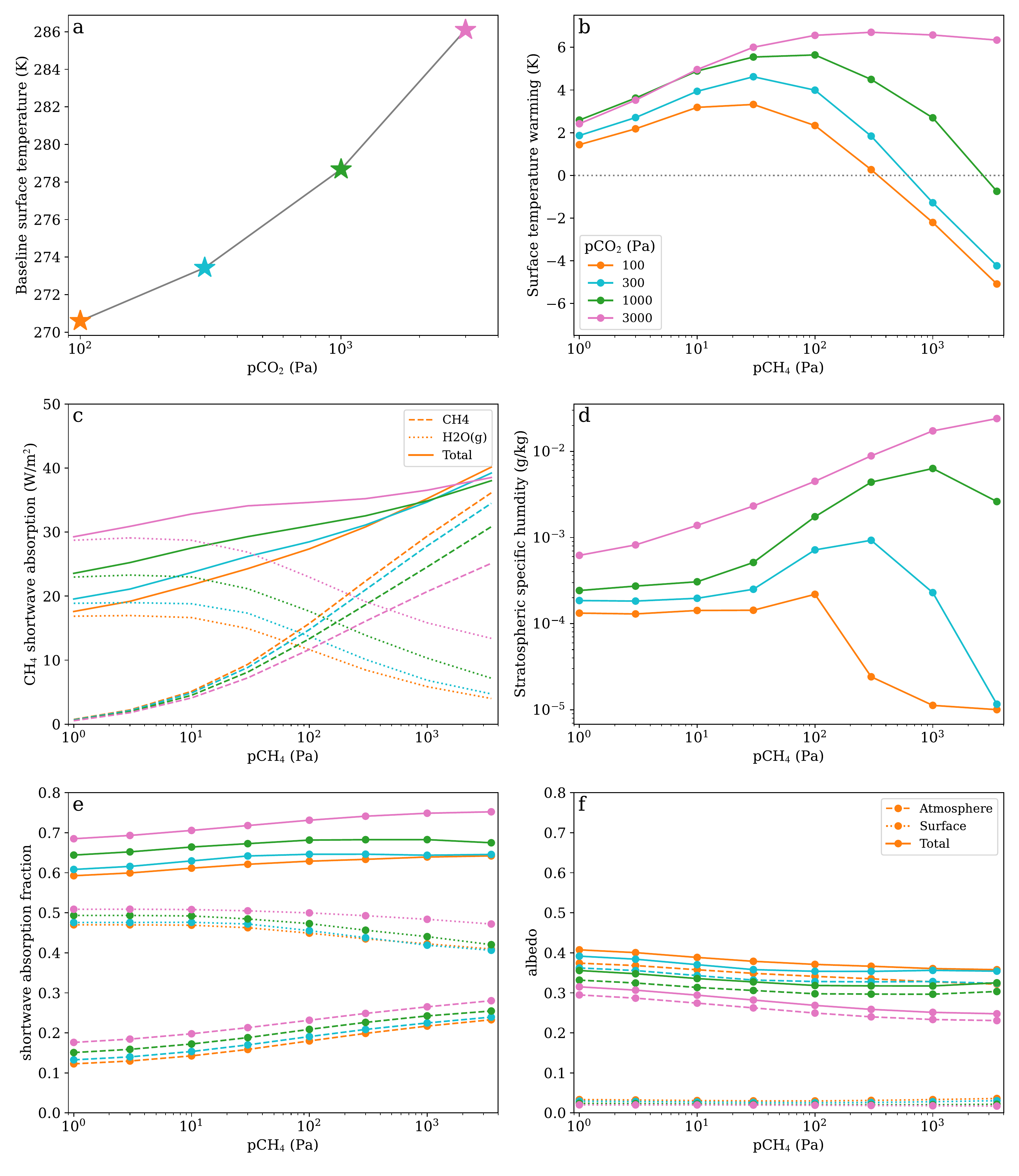}

\caption{(a) shows the global mean surface temperature with respect to the carbon dioxide concentration and assuming p\ch{CH4} is zero, with (b) showing the subsequent change in global mean surface temperature due to the addition of p\ch{CH4} (positive indicates warming). Colours of markers in (a) correspond to the same p\ch{CO2} in (b). Similar plots for near surface specific humidity are shown in \figurename~\ref{fig:q_surf}. (c) show the global average atmospheric absorption of shortwave radiation by methane (dashed) and water vapour (dotted) and their sum (solid). (d) shows the specific humidity in the stratosphere at $\sigma=0.1$. (e) and (f) shows the fractional contribution of atmospheric (dashed), surface (dotted) and total (atmospheric plus surface) to the shortwave radiation budget global mean for shortwave absorption (e) and solar-weighted albedo (f). These were calculated using \citeA{donohoe11}, which accounted for atmospheric attenuation of shortwave radiation reflected by the surface.
\label{fig:T_surf}
}
\end{figure}

We now focus on the global effects of changing p\ch{CH4} and p\ch{CO2}. In this section we demonstrate that methane has a maximum warming potential on the global mean surface temperature of less than 7\,K for p\ch{CO2}$\leq$3,000\,Pa for our 2.7\,Ga Archean-like aquaplanet configuration (see \figurename~\ref{fig:T_surf}b). We compare these results to those of \citeA{byrne15} who used a 1D RCM and find that our GCM simulations predict a stronger cooling effect at high methane concentrations.

The global mean surface temperature increase, caused by the addition of methane, is less than 7\,K. \figurename~\ref{fig:T_surf}a shows the mean surface temperature for our simulated atmospheres without methane, i.e. \ch{N2} and \ch{CO2} only with p\ch{CH4}=0, to act as a baseline for simulations where p\ch{CH4}$>$0. The subsequent temperature change from increasing p\ch{CH4} is shown in \figurename~\ref{fig:T_surf}b. For our simulations the contribution of methane to the greenhouse effect peaks between p\ch{CH4} values of 30 to 300\,Pa, with peak global mean warming ranging from approximately 3.5\,K to nearly 7\,K as shown in \figurename~\ref{fig:T_surf}b. The location and magnitude of the peak in global mean surface temperature is dependent on p\ch{CO2}, with the peak occurring at higher p\ch{CH4} and a larger maximum temperature as p\ch{CO2} is increased. Past the peak, increasing p\ch{CH4} leads to the global mean surface temperature decreasing, for which the decrease is largest for the lowest p\ch{CO2}, with some simulations for p\ch{CO2}~$\leq$~1000\,Pa globally cooler than the simulation with no methane at all (below grey dotted line in \figurename~\ref{fig:T_surf}b), with global mean surface temperature dropping by up to approximately 8\,K at p\ch{CH4}$=3500$\,Pa compared to the maximum methane warming at constant p\ch{CO2} (\figurename~\ref{fig:T_surf}b). This peaked response is caused by the balance between methane's longwave warming effect and its shortwave cooling effect, as discussed in \citeA{byrne15} and presented here in \figurename~\ref{fig:CH4_rad_forcings}a.

Conversely, at the higher p\ch{CO2} amount of 3,000\,Pa, the decrease in surface temperature caused by p\ch{CH4}=3,500\,Pa is minimal. From \figurename~\ref{fig:T_surf}c, it can be seen that the total water vapour and methane shortwave absorption increases less as p\ch{CH4} increases for p\ch{CO2}$=$3,000\,Pa compared to lower values of p\ch{CO2}. This is because the warmer climate at p\ch{CO2}$=$3,000\,Pa allows for more water vapour in the atmosphere, which reduces the methane shortwave absorption due to overlap in their absorption cross sections in the shortwave. At higher p\ch{CO2}, the stratosphere becomes more resistant to the emergence of a stratospheric temperature inversion (see \citeA{byrne15}, \figurename~\ref{fig:T_air_maps} and \figurename~\ref{fig:Tqcloud}), and at p\ch{CO2}$=$3,000\,Pa, a cold trap only forms much deeper in the atmosphere (see \figurename~\ref{fig:Tqcloud}). This allows water vapour to reach higher abundances higher in the atmosphere, with stratospheric water vapour increasing significantly with p\ch{CH4} for the p\ch{CO2}$=$3,000\,Pa case only, shown in \figurename~\ref{fig:T_surf}d. This increase in stratospheric water vapour also acts to enhance the greenhouse effect, and leads to only a marginal drop in the global mean surface temperature for p\ch{CO2}$=$3,000\,Pa in \figurename~\ref{fig:T_surf}b.

Use of a 3D GCM leads to a more significant global cooling at high p\ch{CH4} (see \figurename~\ref{fig:T_surf}b) compared to results from a 1D RCM. p\ch{CO2}=100\,Pa and p\ch{CO2}=1000\,Pa offer the most comparable configurations to \citeA{byrne15}, who used \ch{CO2} abundances of 10$^{-3}$ and 10$^{-2}$. For a \ch{CO2} abundance of 10$^{-2}$, \citeA{byrne15} found a plateau in methane's greenhouse warming with surface temperature remaining constant up to a \ch{CH4} abundance of 10$^{-2}$ from 10$^{-3}$, compared to a $\approx$3\,K drop in average surface temperature from the peak at p\ch{CH4}=100\,Pa for p\ch{CO2}=1000\,Pa at p\ch{CH4}=1000\,Pa (temperature difference for green line between p\ch{CH4}=100\,Pa to p\ch{CH4}=1000\,Pa in \figurename~\ref{fig:T_surf}b). Similarly for a \ch{CO2} abundance of 10$^{-3}$, there is a cooling from the peak of $\approx$1\,K in the 1D model compared to $\approx$5.5\,K for p\ch{CO2}=100\,Pa in our 3D model (temperature difference for orange line between p\ch{CH4}=30\,Pa to p\ch{CH4}=1000\,Pa in \figurename~\ref{fig:T_surf}b). This increases to $\approx$8\,K when p\ch{CH4}=3500\,Pa, however methane was not investigated at such levels in \citeA{byrne15}. These are caused, in part, by the presence of a 3D atmospheric circulation in the GCM, with methane affecting the meridional temperature gradient and Hadley circulation (see Section~\ref{sec:meridional}).

\begin{figure}
\centering
\noindent\includegraphics[width=\textwidth]{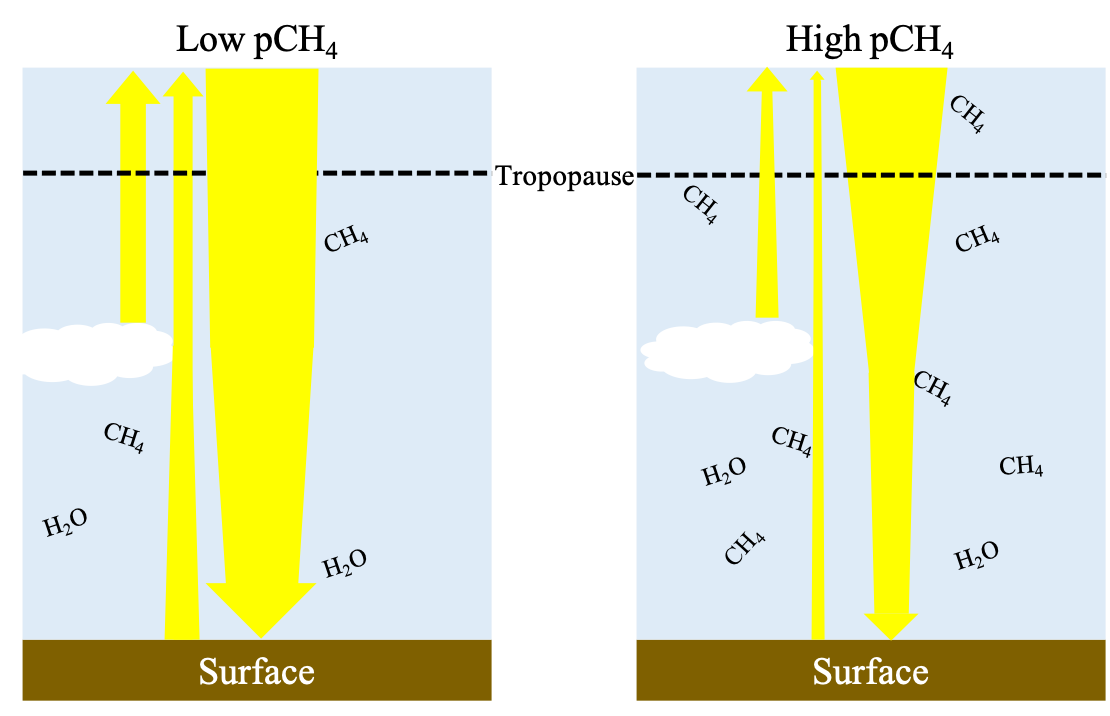}

\caption{Schematic showing how methane affects the vertical distribution of shortwave radiation in the atmosphere for a low (left) and high (right) methane case. Shortwave radiation is represented by yellow arrows. A thinning of the arrow represents a loss of shortwave radiation through absorption in the atmosphere. Upward shortwave radiation represents reflected shortwave radiation from the surface and atmosphere. The tropopause is marked by a dashed line, and for simplicity is placed at the same level in these diagrams.
\label{fig:Schematic}
}
\end{figure}

The addition of methane increases atmospheric shortwave absorption (\figurename~\ref{fig:T_surf}c) in the upper atmosphere and leads to a reduction in planetary albedo. \figurename~\ref{fig:T_surf}e shows this increase in atmospheric absorption of shortwave radiation with p\ch{CH4}. This leads to a decrease in surface shortwave absorption and a reduction in total albedo, shown in \figurename~\ref{fig:T_surf}f. As discussed in Section~\ref{sec:maps} as p\ch{CH4} increases, shortwave heating transitions from predominantly heating the lower troposphere to heating the stratosphere and upper troposphere. This is caused by the increase in shortwave absorption higher in the atmosphere. This is summarised in the schematic in \figurename~\ref{fig:Schematic}.

The increase in shortwave absorption by methane higher in the atmosphere leads to a reduction in the water vapour shortwave absorption, shown in \figurename~\ref{fig:T_surf}c. This is due to spectral overlap between the absorption cross-sections of water vapour and methane in the shortwave \cite<see Figure 6 in>{byrne14}. Increasing p\ch{CH4} leads to an increase in shortwave radiation absorption high in the atmosphere by methane, where specific humidity is lower in concentration compared to lower in the troposphere (see \figurename~\ref{fig:Tqcloud}). However, the relative contributions to shortwave absorption by methane and water vapour also depends on the temperature, with warmer simulations having a higher specific humidity, leading to shortwave absorption higher in the atmosphere. This explains the greater rates of shortwave heating in the troposphere in \figurename~\ref{fig:stream_function} for p\ch{CO2}=3,000\,Pa (compared to p\ch{CO2}=100\,Pa), which is significantly warmer than the other simulations. As a result, these simulations have less shortwave heating by methane at a given p\ch{CH4} (compare magenta and orange dashed lines in \figurename~\ref{fig:T_surf}c).

\subsection{The effect of methane on equator-to-pole temperature gradients}
\label{sec:meridional}

Although the global impacts of methane within our 3D GCM simulations qualitatively match the trends identified in \citeA{byrne15}, differences are apparent between the 1D and 3D results. In this section we describe the impacts of changing meridional transport across our simulations, which cannot be captured in 1D models. We find that the equator-to-pole temperature difference tends to increase with p\ch{CH4} initially due to an increase in methane radiative forcing at the equator relative to the poles. As p\ch{CH4} increases further, equator-to-pole temperature differences plateau due to an increase in radiative forcing at the poles and a decrease in efficiency in meridional heat transport.

\begin{figure}
\centering
\noindent\includegraphics[width=\textwidth]{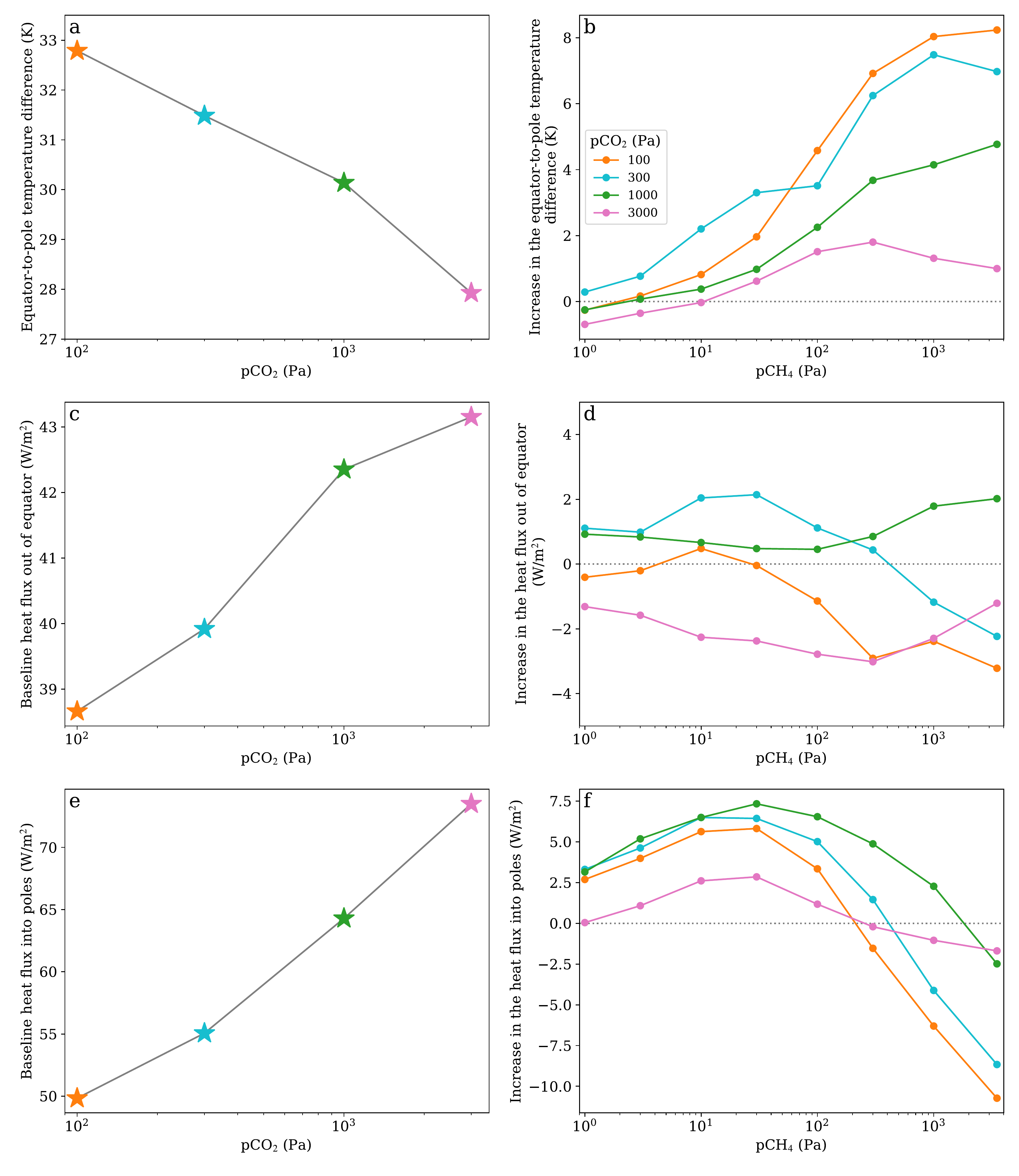}

\caption{(a) shows the average equator-to-pole temperature difference with only carbon dioxide (p\ch{CH4}=0), with (b) showing the subsequent increase in equator-to-pole temperature difference due to the addition of methane (positive indicates an increase in the equator-to-pole temperature difference). (c) shows the atmospheric heat flux out of the equatorial region for different p\ch{CO2} with p\ch{CH4}=0 (referred to as baseline cases), while (d) shows the increase in the equatorial heat flux from the subsequent addition of methane to (c). (e) shows the atmospheric heat flux into the polar regions for different p\ch{CO2} (with p\ch{CH4}=0), while (f) shows the increase in the polar heat flux from the subsequent addition of methane to (e). Calculations for (c)-(f) use methods from \citeA{lambert11} to calculate atmospheric heat fluxes (see \ref{sec:lambert11} for more detail). The equatorial region is considered as spanning latitudes of 10\degree\,S to 10\degree\,N, while the polar regions are 70\degree\,S/N to the pole. Colours of data points in (a), (c) and (d) correspond to the same p\ch{CO2} in (b), (d) and (f).
\label{fig:EquPol_Tdiff}
}
\end{figure}

The equator-to-pole temperature gradient reduces with increasing p\ch{CO2}, demonstrated in \figurename~\ref{fig:EquPol_Tdiff}a. This is due to an increase in the atmospheric temperature which increases the moist static energy transport \cite{Kaspi15}. The meridional stream functions also increases in latitudinal and vertical extent, as shown in \figurename~\ref{fig:stream_function}. However, the meridional temperature gradient's response to methane is more complex.

\figurename~\ref{fig:EquPol_Tdiff}b shows the increase in equator-to-pole temperature difference as a result of the addition of methane to the simulations. From \figurename~\ref{fig:EquPol_Tdiff}b we find that increasing p\ch{CH4} generally causes an increase in equator-to-pole temperature differences for p\ch{CH4}$\leq$300\,Pa, above this the equator-to-pole temperature difference either begins to plateau or decreases. The exact change is dependent on the p\ch{CO2}, with the behaviour more extreme for lower p\ch{CO2}, but the general trends remain. As well as this the main drivers for the meridional temperature gradients are methane's varying meridional radiative forcing (\figurename~\ref{fig:CH4_rad_forcings}b,c) and the effect on the meridional heat transport.

\begin{figure}
\centering
\noindent\includegraphics[width=\textwidth]{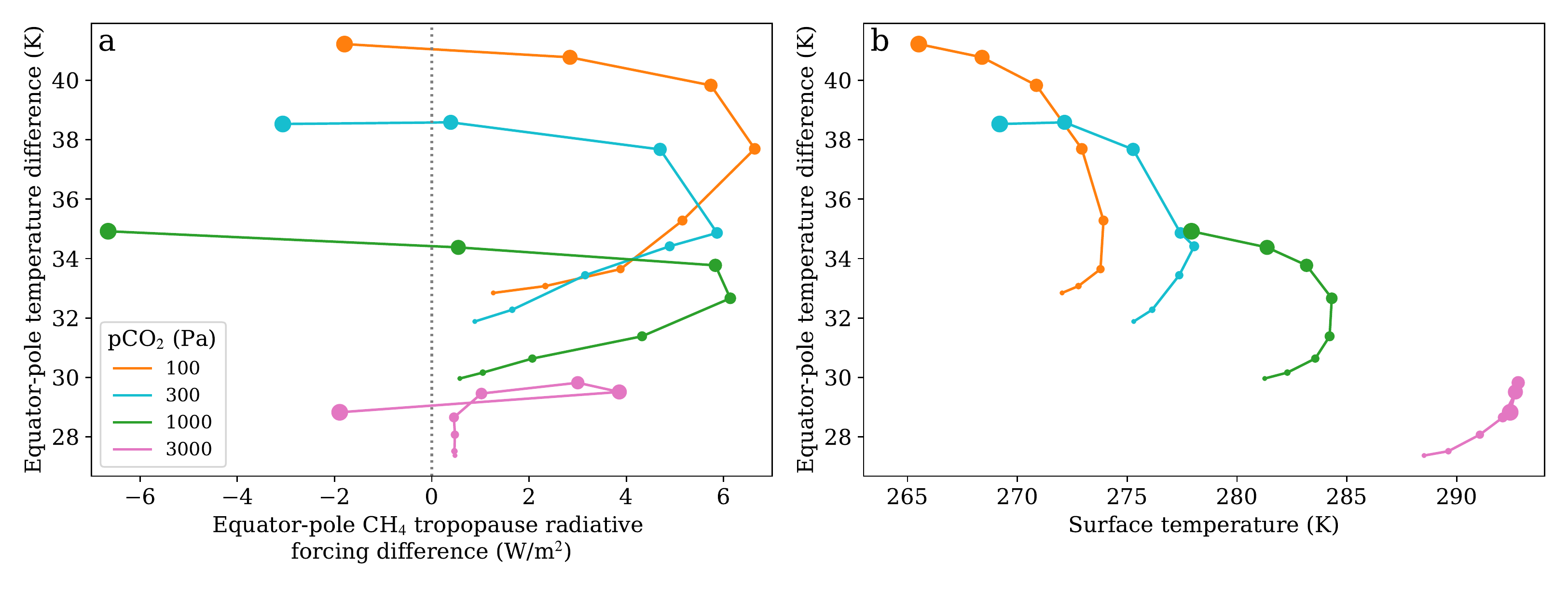}

\caption{(a) Shows the equator-to-pole temperature difference against the difference in equatorial and polar net downward p\ch{CH4} tropopause radiative forcing (positive indicates an increase in radiative forcing at the equator relative to the poles). (b) shows the global mean surface temperature against the equator-to-pole temperature difference. The equatorial region is considered as spanning latitudes of 10\degree\,S to 10\degree\,N, while the polar regions are 70\degree\,S/N to the pole. p\ch{CH4} is represented through the size of the marker (larger maker equivalent to a higher p\ch{CH4}) for p\ch{CH4} values of $\{1,3,10,30,100,300,1000,3500\}$\,Pa.
\label{fig:meridional_heating}
}
\end{figure}

\figurename~\ref{fig:meridional_heating}a shows that for p\ch{CO2}$\leq$1000\,Pa increasing p\ch{CH4} initially leads to an increase in the relative \ch{CH4} tropopause forcing at the equator compared to the pole, which drives an increase in the equator to pole temperature difference. This suggests, that for p\ch{CH4}$\leq$300\,Pa, the equator-to-pole temperature difference in \figurename~\ref{fig:EquPol_Tdiff}b is driven by the differences in \ch{CH4} radiative forcing. 

The plateau that emerges in the equator-to-pole temerature differences in \figurename~\ref{fig:EquPol_Tdiff}b as p\ch{CH4} increases above 300\,Pa, can be explained by a balance between a weakening of the meridional heat transport and an increase in methane radiative forcing at the poles relative to the equator. For p\ch{CO2}$\leq$1000\,Pa, as p\ch{CH4} increases above 100\,Pa, the difference in the equator-to-pole radiative forcing decreases in \figurename~\ref{fig:meridional_heating}a, with polar forcing approaching equality to the equatorial radiative forcing and eventually exceeding it at p\ch{CH4}$=$3500\,Pa. However the change in equator-to-pole temperature differences over this period shows only a small (order of 2\,K between 300-3500\,Pa) generally increasing trend compared to the large increase in the relative methane radiative forcing between the equator and pole (\figurename~\ref{fig:meridional_heating}a), which would work to decrease the equator-to-pole temperature difference. The minimal change in the equator-to-pole temperature difference is caused by a reduction in the meridional heat transport, which offsets the increase in methane radiative forcing at the poles relative to the equator. Meridional heat transport weakens due to a reduction in the latitudinal extent of the Hadley circulation (\figurename~\ref{fig:stream_function}) and a reduction in global mean surface temperature compared to the peak  in \figurename~\ref{fig:T_surf}b, which would act to increase equator-to-pole temperature differences (see \figurename~\ref{fig:meridional_heating}b). This balance maintains a relatively small change in equator-to-pole temperature differences for p\ch{CH4}$\geq$300\,Pa in \figurename~\ref{fig:EquPol_Tdiff}b.

At p\ch{CO2}=3,000\,Pa, the equator-to-pole temperature difference change is smaller compared to the other case, changing by $\approx$6\,K in total from p\ch{CH4}$=$1\,Pa to p\ch{CH4}$=$3,500\,Pa. This is partly due to the latitudinal extent of the Hadley circulation in \figurename~\ref{fig:stream_function} remaining similar at p\ch{CO2}=3,000\,Pa, which may be more important in controlling equator-to-pole temperature difference compared to the change in radiative forcing between the equator and poles. For p\ch{CH4}$\leq$30\,Pa and p\ch{CO2}=3,000\,Pa relative radiative forcing between the equator and poles remains constant (\figurename~\ref{fig:meridional_heating}a). At p\ch{CO2}=3,000\,Pa for p\ch{CH4}$>$30\,Pa equatorial forcing increases relative to the poles, with the equator-to-pole temperature difference increasing. The reduction in temperature difference between p\ch{CH4}=1,000\,Pa and p\ch{CH4}=3,500\,Pa may be partially driven by larger methane radiative forcing at the poles compared to the equator at p\ch{CH4}$=$3,500\,Pa.

A reduction in the meridional heat transport with increasing p\ch{CH4} above 30\,Pa contributes towards an increase in the equator-to-pole temperature gradient. As discussed in Section~\ref{sec:maps} the Hadley circulation reduces in vertical and latitudinal extent when increasing p\ch{CH4} (see \figurename~\ref{fig:stream_function}). The effect this has on the atmospheric heat transport for the equatorial and polar regions is shown in \figurename{s}~\ref{fig:EquPol_Tdiff}c-f. We used the net flux difference between the surface and top of atmosphere to estimate the flux transported horizontally into a region, as described in \citeA{lambert11} (see \ref{sec:lambert11} for more detail), and shown for our simulations in \figurename{s}~\ref{fig:EquPol_Tdiff}c-f. The change in heat flux out of the equator is relatively small compared to the poles, with changes on order of less than 5$\%$ due to changes in methane (by comparing \figurename~\ref{fig:EquPol_Tdiff}c and \figurename~\ref{fig:EquPol_Tdiff}d). \figurename~\ref{fig:EquPol_Tdiff}f shows that as p\ch{CH4} increases, there is an initial increase in heat flux into the poles, which peaks at p\ch{CH4}$\approx$30\,Pa, potentially a result of the warmer global temperatures increasing the energy transported into the poles through moisture. Above p\ch{CH4}$\approx$30\,Pa heat flux into poles reduces again, which is expected from the weakening circulation in \figurename~\ref{fig:stream_function}, and a corresponding sharp decrease in polar heat flux can be seen with a corresponding sharp rise in the equator-to-pole temperature difference (shown in \figurename~\ref{fig:EquPol_Tdiff}b). The equator-to-pole temperature difference's response to increasing p\ch{CH4} is also dependent on p\ch{CO2}, as shown in \figurename~\ref{fig:EquPol_Tdiff}b.

\begin{figure}
\centering
 \noindent\includegraphics[width=0.5\textwidth]{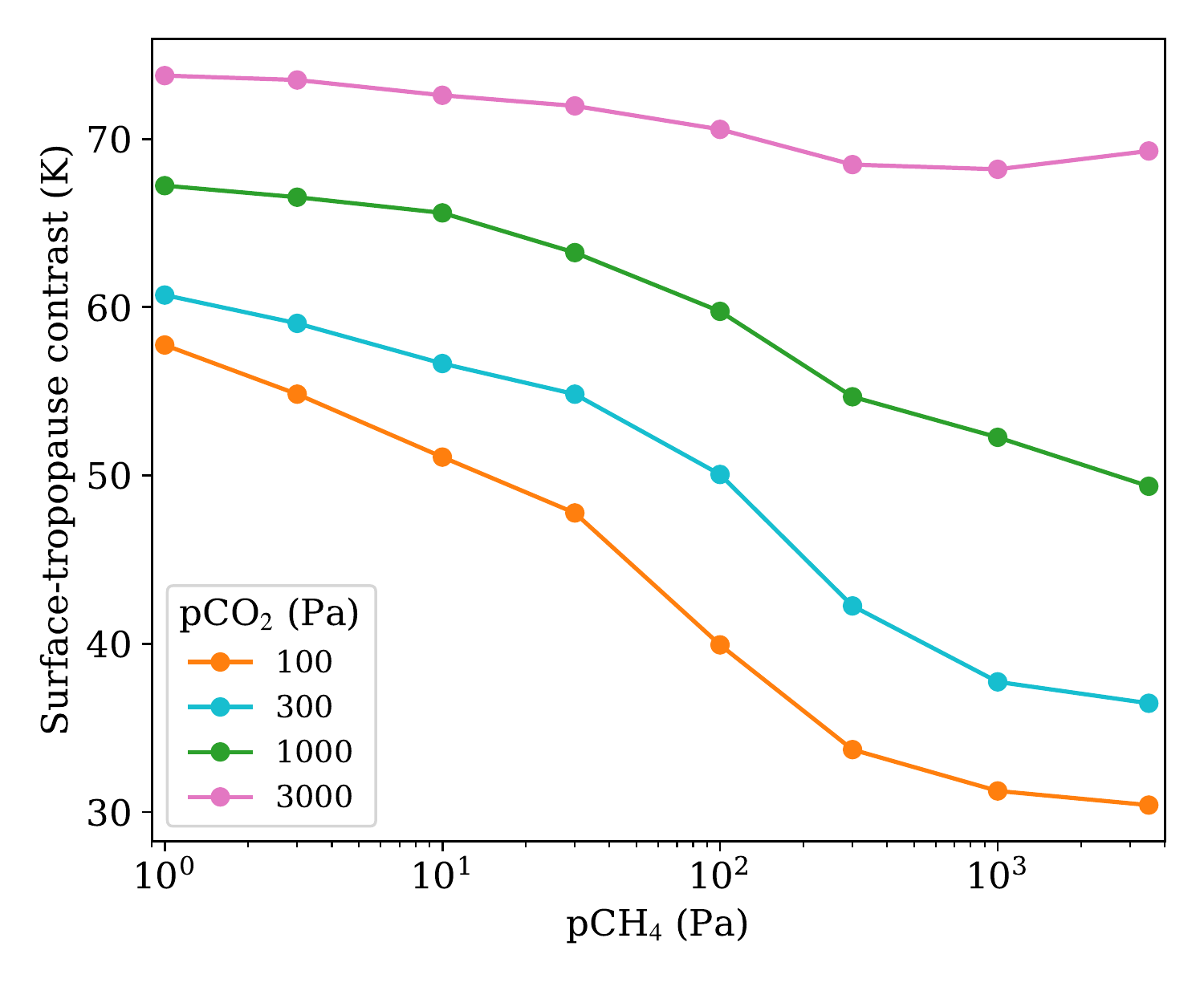}

\caption{Shows the temperature contrast between the surface and atmosphere at $\sigma$=0.2 horizontally averaged over the polar regions. The poles span from 70\degree\,N/S to the pole. $\sigma$=0.2 is used to approximate the tropopause height.
\label{fig:T_surf_atmos_contrast}
}
\end{figure}

The enhanced longwave forcing by methane at higher p\ch{CO2} values in \figurename~\ref{fig:CH4_rad_forcings} is due to an increasing temperature contrast between the surface and upper troposphere. \figurename~\ref{fig:T_surf_atmos_contrast} shows this change in temperature contrast for the polar regions. The temperature contrast increases with p\ch{CO2}, leading to the increased longwave forcing seen in \figurename~\ref{fig:CH4_rad_forcings}a-c. In contrast, methane acts to warm the stratosphere and upper troposphere by absorbing solar radiation and thus increasing methane decreases the temperature contrast, shown in \figurename~\ref{fig:T_surf_atmos_contrast}.

\begin{figure}
\centering
\noindent\includegraphics[width=\textwidth]{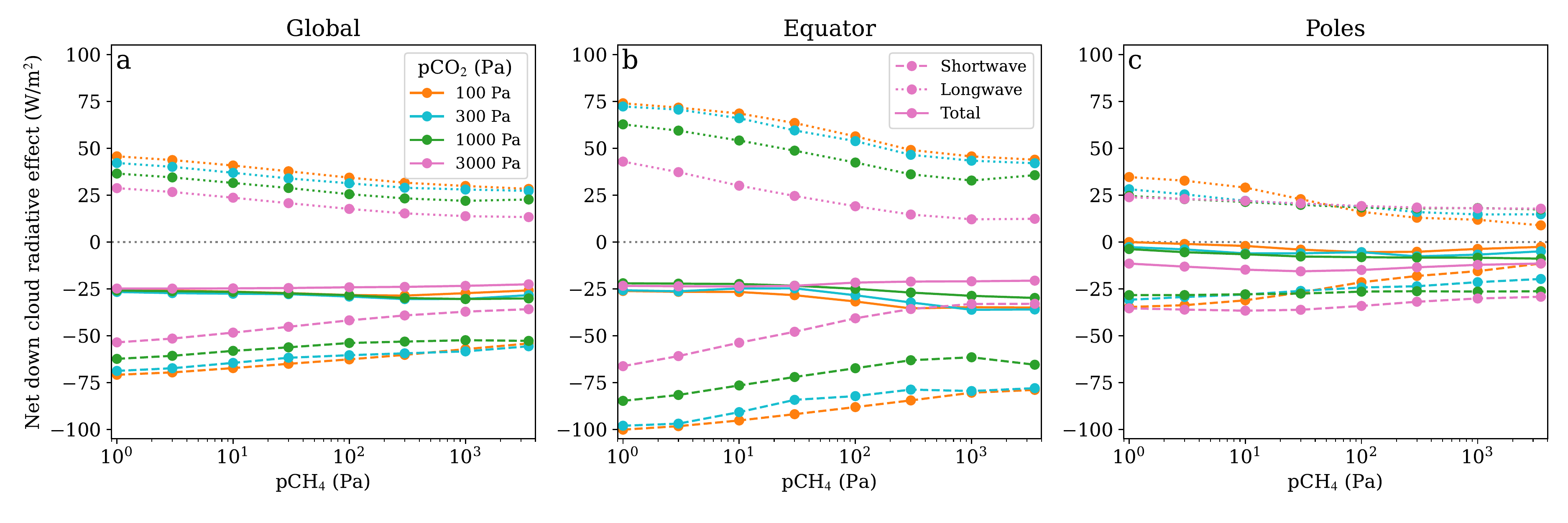}

\caption{(a) shows the global average net down cloud radiative effect at $\sigma$=0.2 (approximate location of the tropopause) where $\sigma$ is the pressure divided by the global mean surface pressure. Cloud radiative effect is shown for shortwave (dashed), longwave (dotted) and their sum (solid). (b) and (c) show the same as (a) averaged over 10$\degree$S to 10$\degree$N as the equator in (b), and poleward of 70$\degree$N/S for the polar regions in (c).
\label{fig:cloud_rad_forcings}
}
\end{figure}

For this study clouds play only a secondary effect, and are not, therefore, discussed in detail. \figurename~\ref{fig:cloud_rad_forcings} shows the cloud radiative effect at the tropopause, and although the magnitude of the change in longwave and shortwave cloud radiative effect vary the total cloud radiative effect remains nearly constant. Cloud fractions horizontally averaged pressure-temperature plots for the equator and poles are shown in \figurename~\ref{fig:Tqcloud}. We are currently working on a more in-depth study of the impacts of clouds over this parameter space.


\section{Discussion}
\label{sec:discussion}

GCMs provide useful insight into the role of methane in the Archean climate, both validating effects found in RCMs and also finding important differences that come from 3D modelling. The initial global warming due to methane is similar between both models, with the models diverging when predicting temperatures for p\ch{CH4}:p\ch{CO2} $>$ 0.1. In this range GCMs predict large temperature drops, particularly for cooler, lower \ch{CO2} configurations, with temperatures dropping by up to 8\,K from the peak temperature.

GCMs are also useful in their ability to investigate the meridional heat distribution, where methane plays three important roles. 1) Differences in equator-to-pole methane radiative forcing: equator dominated for p\ch{CH4}:p\ch{CO2} $<$ 0.1, and transitioning to polar dominated above this. 2) Hadley circulation weakens when increasing p\ch{CH4}, which reduces meridional heat transport. 3) Atmospheric temperatures affect the moisture held in the atmosphere, so the peaking in temperature at p\ch{CH4}:p\ch{CO2} $=$ 0.1 leads to increases in meridional heat transport below this and decreasing heat transport above this when increasing p\ch{CH4}. It may be possible to replicate this in a 2 column model (equatorial and polar column) by parameterising these three factors.

Our study supports that the FYS paradox can be resolved based on p\ch{CO2} constraints around 2.7\,Ga. Global mean surface temperatures close to the pre-industrial Earth, can be achieved with p\ch{CO2}$=$3000\,Pa and p\ch{CH4}$=$0\,Pa and p\ch{CO2}$=$1000\,Pa and p\ch{CH4}$\approx$100\,Pa (\figurename~\ref{fig:T_surf}a,b). The first of these falls within the range of p\ch{CO2} values constrained by \citeA{kanzaki15} from 2.75\,Ga, while the latter composition falls within the constraint of \citeA{driese11} at 2.7\,Ga. The addition of methane to p\ch{CO2}$=$3000\,Pa up to p\ch{CH4}$=$3500\,Pa would lead to global mean surface temperatures up to 7\,K warmer than the pre-industrial Earth, while increasing or decreasing methane from the p\ch{CO2}$=$1000\,Pa and p\ch{CH4}$\approx$100\,Pa configuration would lead to global mean temperatures decreasing. For p\ch{CO2}$=$300\,Pa and p\ch{CO2}$=$100\,Pa, global mean temperatures could reach up to 278 and 274\,K respectively, with optimal methane abundances. In our simulations, these remain ice free, with \citeA{wolf13} finding that these concentrations of \ch{CO2} with some \ch{CH4} also avoid being fully glaciated, with at least 50\% of the surface remaining ice free. Higher p\ch{CO2} values than 3000\,Pa are not necessary for surface pressures around 10$^5$\,Pa, but could be important to warm the Archean Earth if the total surface pressure was lower.

Some studies have suggested that surface pressures may have been lower than the present day. Fossil raindrops have been used to approximate surface pressure at 2.7\,Ga to between 5$\times 10^4$ to 1.1$\times 10^5$\,Pa \cite{som12}, although debate remains over the upper limit of this constraint \cite{kavanagh15}. Furthermore, fossilised gas bubbles in lava flows at 2.74\,Ga predict a surface pressure of 2.3$\pm$2.3$\times 10^4$\,Pa \cite{som16}. The understanding of the effect of methane on the climate at different pressures is thus very important for the Archean, to be considered in future studies.

Methane has a secondary impact on the climate through its importance in the formation of hydrocarbon hazes, and it is suggested that there may have been periods during the Archean where a haze layer was present \cite{domagalgoldman08,izon15}. Haze forms when the \ch{CH4}:\ch{CO2} ratio exceeds approximately 0.1 \cite{trainer06}, and potentially has a cooling effect on climate by up to 20\,K \cite{arney16}. \citeA{arney16} found that a thick haze was produced for p\ch{CH4}:p\ch{CO2}$>$0.2. For 0.1$<$p\ch{CH4}:p\ch{CO2}$<$0.2 a thin haze is produced that is spectrally indistinguishable from an atmosphere with no haze \cite{arney16}. The peak surface temperature in our experiments were found to occur around p\ch{CH4}/p\ch{CO2}$>$0.1 so we expect this peak to remain with the presence of haze, and for p\ch{CH4}/p\ch{CO2}$>$0.2 hazes form and accelerate cooling further. Although we intend to study hazes in future work, with global mean surface temperatures for a given p\ch{CO2} tending to peak around p\ch{CH4}=0.1$\times$p\ch{CO2}, higher methane concentrations are not able to increases global mean surface temperatures further, even without considering the effects of haze.

\citeA{byrom22} and \citeA{collins18} found that surface albedo can affect the shortwave radiative forcing, with higher albedos at wavelengths where methane absorbs strongly leading to an increase in absorption of upward radiation leading to a positive radiative forcing, which can arise from high albedo on land surfaces such as deserts \cite{byrom22}. Land fraction remains uncertain for the Archean \cite<see discussion in>{charnay20}, with a general consensus that the fraction of continental crust was less than present day \cite{hawkesworth19}, as well as the potential for a larger ocean volume \cite{dong21}, with suggestions land fraction could have been as low as 2-3\% until 2.5\,Ga \cite{flament08}. We test the sensitivity of the tropopause radiative forcing by increasing global ocean albedo to 0.533 (value for albedo of quartz sand in near infrared, used in \citeA{byrom22}) and run the model for one time step and compare the radiative forcings, shown in \ref{sec:RF_albedo}. Using this extreme value for albedo, we find that the shortwave radiative forcing becomes less negative. However the general trend for radiative forcing remains the same. Thus, for a surface with higher albedo, we would expect the decline in global mean surface temperatures to be less at higher methane values. Further exploration is required to explore the effect of different continent configurations on the methane radiative forcing and the overall effect on climate, with a spectrally dependent surface albedo.

Although methane, across the range we have studied, has a relatively small effect on global mean surface temperatures of --6 to +7\,K, its abundance has a more significant impact on the meridional heat transport, which could affect the likelihood of the planet entering a snowball state. High methane levels generally increase the equator-to-pole difference by up to 8\,K, which combined with a realistic ice-albedo feedback could affect the stability of the temperate state -- which in this case refers to a state that is not fully glaciated. However full understanding of this would require consideration of ocean heat transport, continental configuration and a thermodynamic treatment of sea-ice, all of which are beyond the scope of this study, which focuses on the climate differences stemming from methane's radiative effect.

Theoretical opacity data that further fill in missing regions of methane's shortwave absorption up to 833\,nm \cite{yurchenko14} and 746\,nm \cite{hargreaves20} have also become available that allows for further investigation into the significance of methane's shortwave absorption. However, line lists for methane covering much of the wavelength region covered by the Sun's emission are not yet available, motivating the extension of work generating line lists to cover this missing absorption data.

Potentially habitable planets have now been detected orbiting M-dwarf stars, which are cooler than the Sun. For these exoplanets the shortwave absorption would be stronger due to a larger fraction of the stellar emission coming from the near infrared, for M-dwarfs compared to G-dwarfs like the Sun, where methane has strong absorption. The impacts of missing absorption data, as discussed above, become even more important due to the host star emitting more radiation at longer wavelengths where absorption from greenhouse gases is higher \cite{eager20}. If these planets host life, their early environment may have resembled that of the Archean Earth, given that a significant fraction of Earth's own `inhabited' history is represented by this period \cite{catling20}. Therefore, these planets could have high concentrations of methane and an even stronger temperature sensitivity to methane abundance due to the interaction of the host star spectrum and the methane rich atmosphere.

\section{Conclusions}
\label{sec:conclusions}

We used the Met Office Unified Model to understand the potential effect of atmospheric \ch{CH4} and \ch{CO2} on the climate of the Archean at 2.7\,Ga. A productive biosphere could have led to an atmosphere with p\ch{CH4} ranging from 10--3,500\,Pa \cite{kharecha05}. We found that the greenhouse effect of methane does not exceed a global average of 7\,K for p\ch{CO2}$\leq$3,000\,Pa, with the peak temperature occurring for p\ch{CH4} between 30--300\,Pa. This peak in temperature shifts to higher p\ch{CH4} for higher values of p\ch{CO2}. The peaked effect is due to methane's absorption of shortwave radiation in the stratosphere that reduces heating of the surface, which dominates over the greenhouse effect of methane at higher p\ch{CH4}. We confirm results from \citeA{byrne15} and extend them to 3D simulations, but also demonstrate that use of a 3D GCM is important to fully capture cooling of the climate at high methane concentrations, due to changes in meridional circulation and methane radiative forcing. Maximum temperatures are reached when the \ch{CH4}:\ch{CO2} is approximately 0.1, below the ratio where significant cooling occurs due to the presence of haze \cite{arney16}. More work is required to understand the interaction of biogeochemical cycles with the atmospheric composition and potential haze formation. 

\section*{Open Research}

The research data supporting this publication are openly available from the University of Exeter's institutional repository at: \url{https://doi.org/10.24378/exe.4347} with  CC BY 4.0 \cite{eagernash22}.

\appendix
\section{Additional figures}
\label{sec:appendix}
\begin{figure}
\centering
 \makebox[\textwidth][c]{\noindent\includegraphics[width=1.2\textwidth]{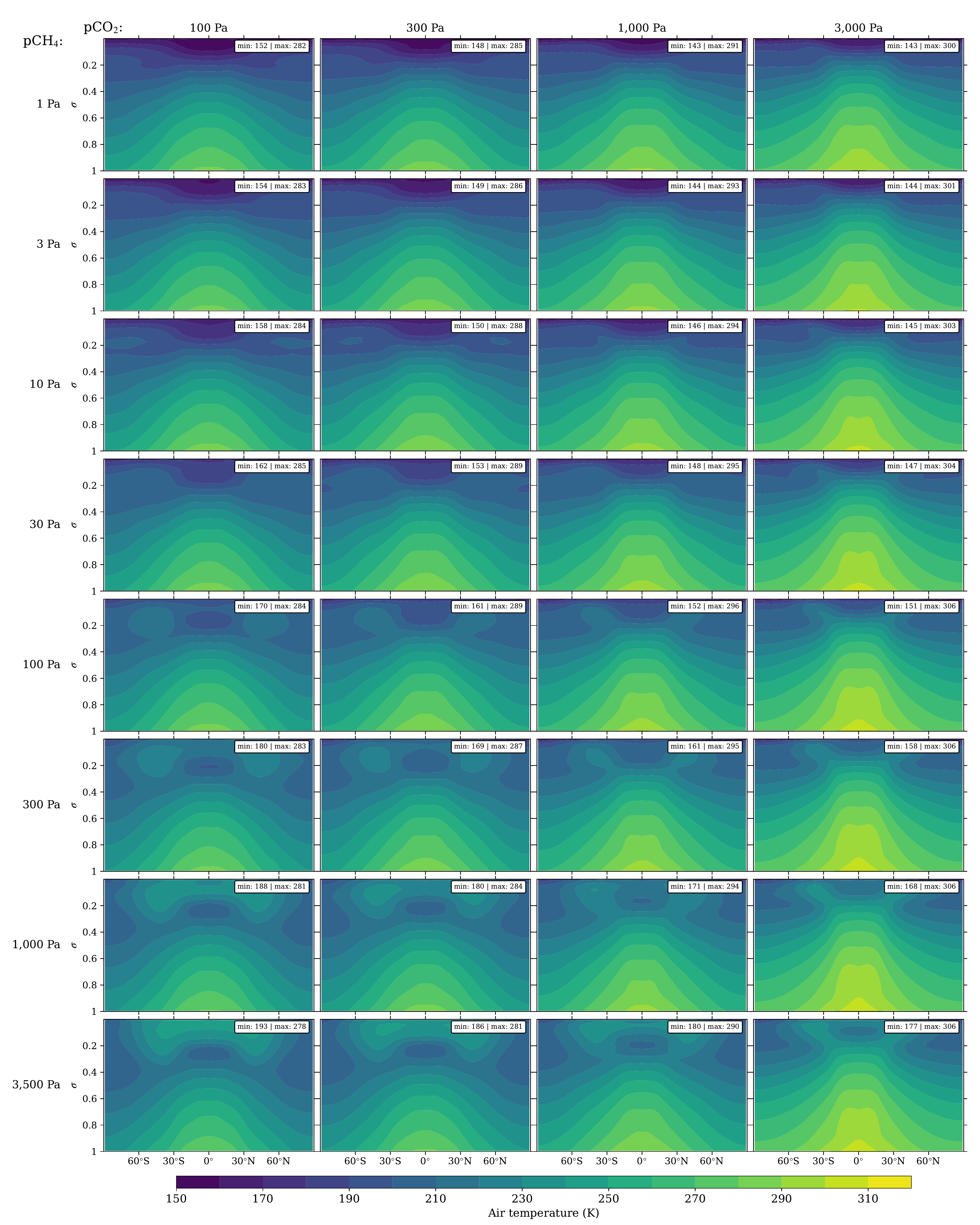}}

\caption{Zonal averaged air temperature (colour scale), for increasing surface partial pressures of carbon dioxide from left to right, and methane from top to bottom. Plotted as latitude vs $\sigma$, where $\sigma$ is the pressure divided by the global mean surface pressure. The same colour scale is used for each plot, with maximum and minimum temperatures for each simulation displayed in the top right of each sub figure.
\label{fig:T_air_maps_full_grid}
}
\end{figure}

\begin{figure}
\centering
 \makebox[\textwidth][c]{\noindent\includegraphics[width=1.2\textwidth]{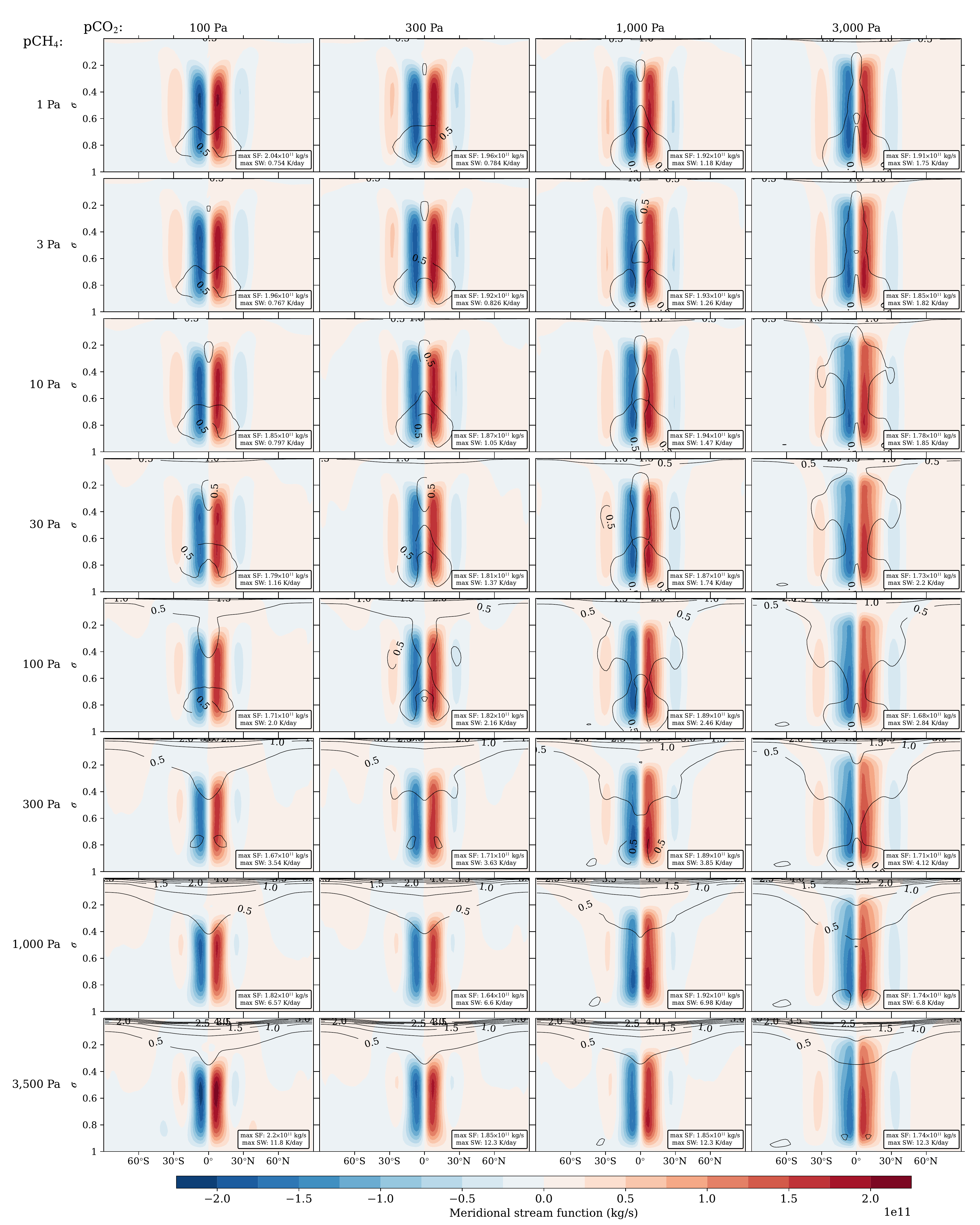}}

\caption{Zonal averaged meridional stream functions (colour scale), for increasing surface partial pressures of carbon dioxide from left to right, and methane from top to bottom. Positive and negative values represent clockwise and anticlockwise circulation respectively. Contours show the heating of the atmosphere due to shortwave radiation in K/day. The same colour scale is used for each plot, with maximum values for the stream function (SF) and shortwave heating rate (SW) are shown for each simulation in the bottom right of each sub figure.
\label{fig:HC_full}
}
\end{figure}

\begin{figure}
\centering
\noindent\includegraphics[width=\textwidth]{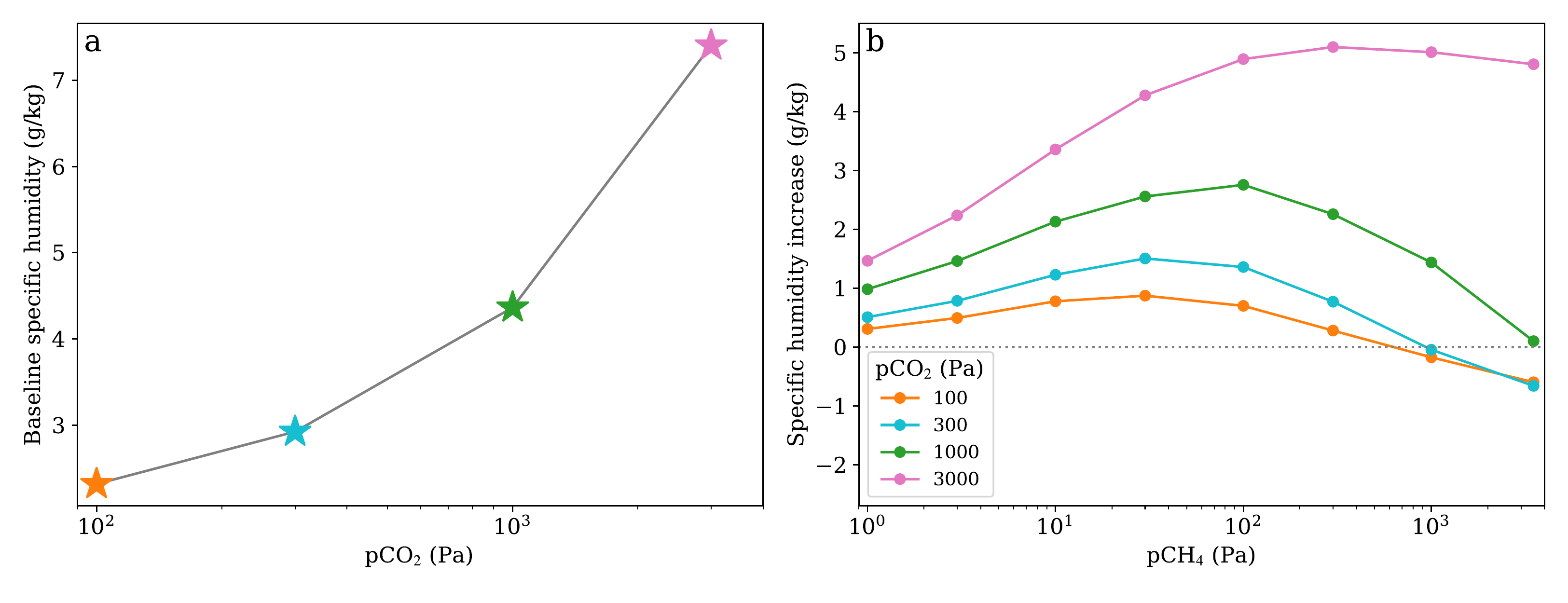}

\caption{Similar to \figurename~\ref{fig:T_surf} (a) shows the global mean near surface ($\sigma=1.0$) specific humidity with respect to the carbon dioxide concentration and assuming p\ch{CH4} is zero, with (b) showing the subsequent change in global mean near surface specific humidity due to the addition of p\ch{CH4} (positive indicates warming). Colours of markers in (a) correspond to the same p\ch{CO2} in (b).
\label{fig:q_surf}
}
\end{figure}

\begin{figure}
\centering
 \makebox[\textwidth][c]{\noindent\includegraphics[width=0.9\textwidth]{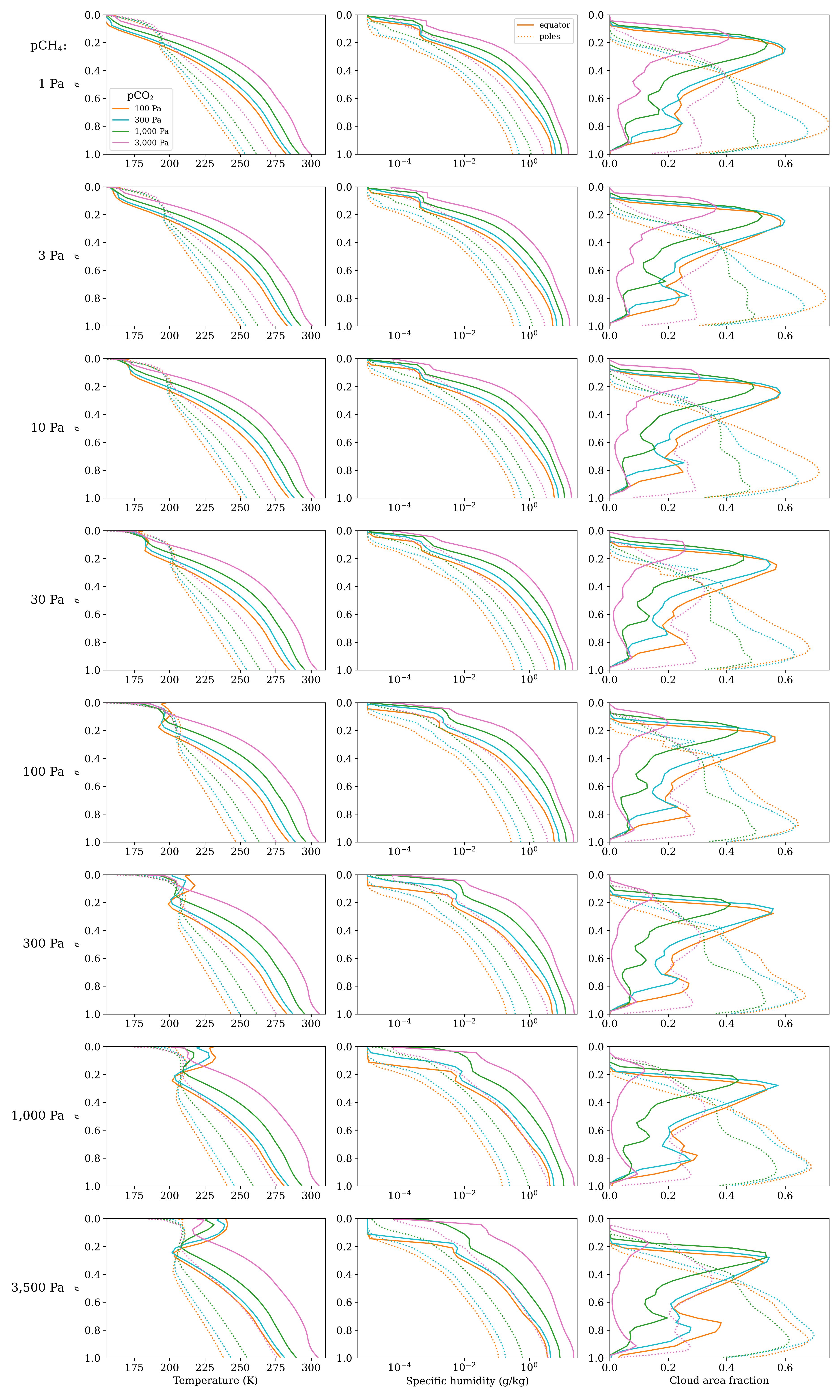}}

\caption{Shows horizontally averaged air temperature (left column), specific humidity (middle column) and cloud area fraction (right column) for the equator (solid lines) and poles (dotted lines), for increasing surface partial pressures of methane from top to bottom. The equator is considered as spanning latitudes of 10\degree\,S to 10\degree\,N, while the poles are 70\degree\,S/N to the pole.
\label{fig:Tqcloud}
}
\end{figure}

\begin{figure}
\centering
 \makebox[\textwidth][c]{\noindent\includegraphics[width=\textwidth]{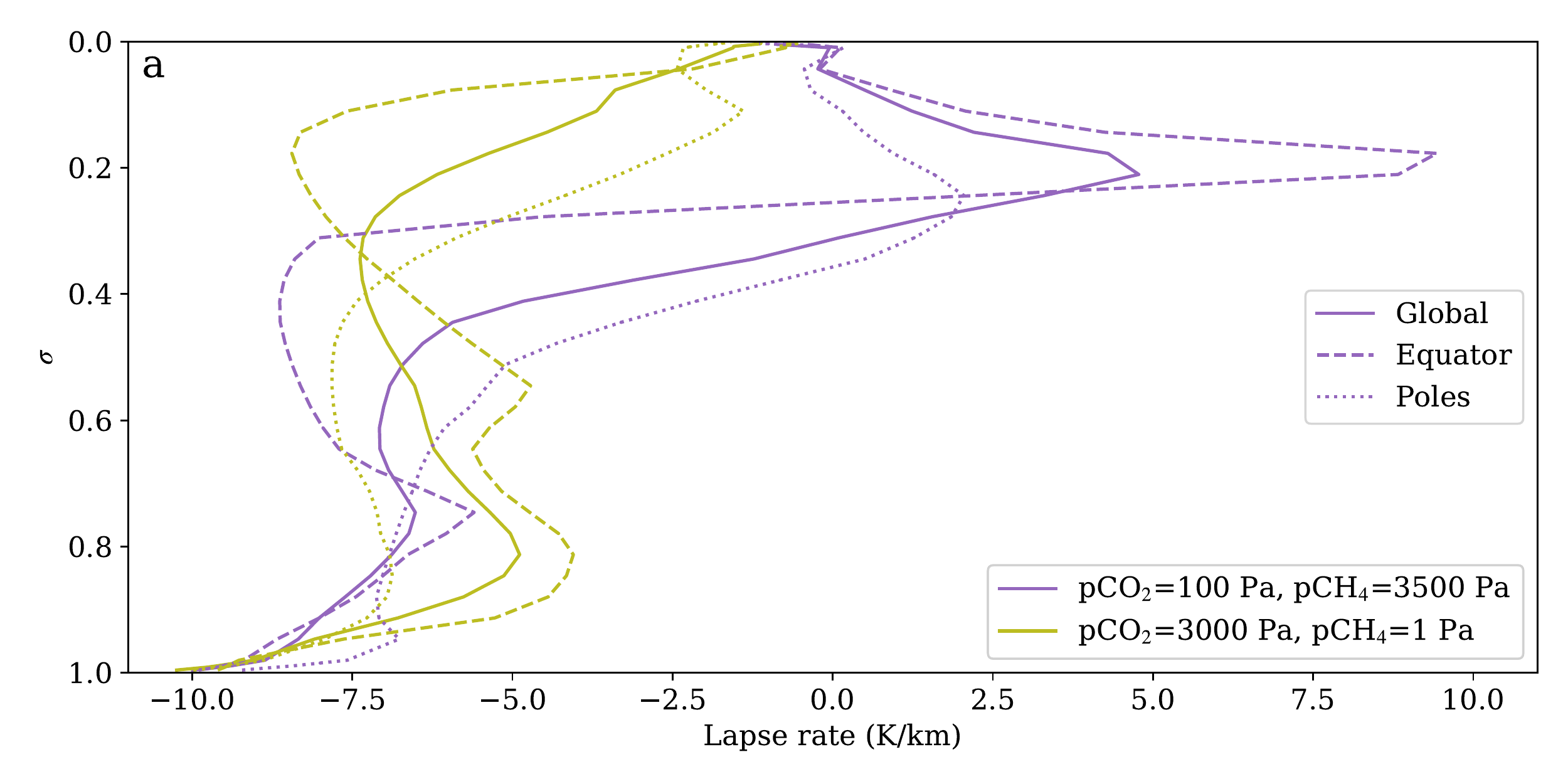}}

\caption{Shows lapse rate for the global (solid lines), equatorial (dashed lines) and polar (dotted lines) averages, for p\ch{CO2}=100\,Pa and p\ch{CH4}=3500\,Pa (purple) and p\ch{CO2}=3000\,Pa and p\ch{CH4}=1\,Pa (green). The equator is considered as spanning latitudes of 10\degree\,S to 10\degree\,N, while the poles are 70\degree\,S/N to the pole.
\label{fig:lapse_rate}
}
\end{figure}

Here, we present results from all of our simulations for cases matching figures in the main paper, as well as additional figures relating to specific humidity, clouds and lapse rate. \figurename~\ref{fig:T_air_maps_full_grid} and \figurename~\ref{fig:HC_full} shows the same information as \figurename~\ref{fig:T_air_maps} and \figurename~\ref{fig:stream_function} respectively, but for the full grid of simulations. \figurename~\ref{fig:q_surf} shows the global mean near surface specific humidity change for different p\ch{CO2} and p\ch{CH4} values, similar to \figurename~\ref{fig:T_surf}a,b. \figurename~\ref{fig:Tqcloud} show air temperature, specific humidity and cloud area fractions to supplement data presented in  \figurename~\ref{fig:T_surf}, \figurename~\ref{fig:CH4_rad_forcings} and \figurename~\ref{fig:T_surf_atmos_contrast}. \figurename~\ref{fig:lapse_rate} shows the lapse rate for a low \ch{CO2} and high \ch{CH4} case (p\ch{CO2}=100\,Pa, p\ch{CH4}=3500\,Pa) and a high \ch{CO2} and low \ch{CH4} case (p\ch{CO2}=3000\,Pa, p\ch{CH4}=1\,Pa).


\section{Spectral resolution sensitivity analysis}
\label{sec:spec_file_test}

\begin{figure}
\centering
 \makebox[\textwidth][c]{\noindent\includegraphics[width=\textwidth]{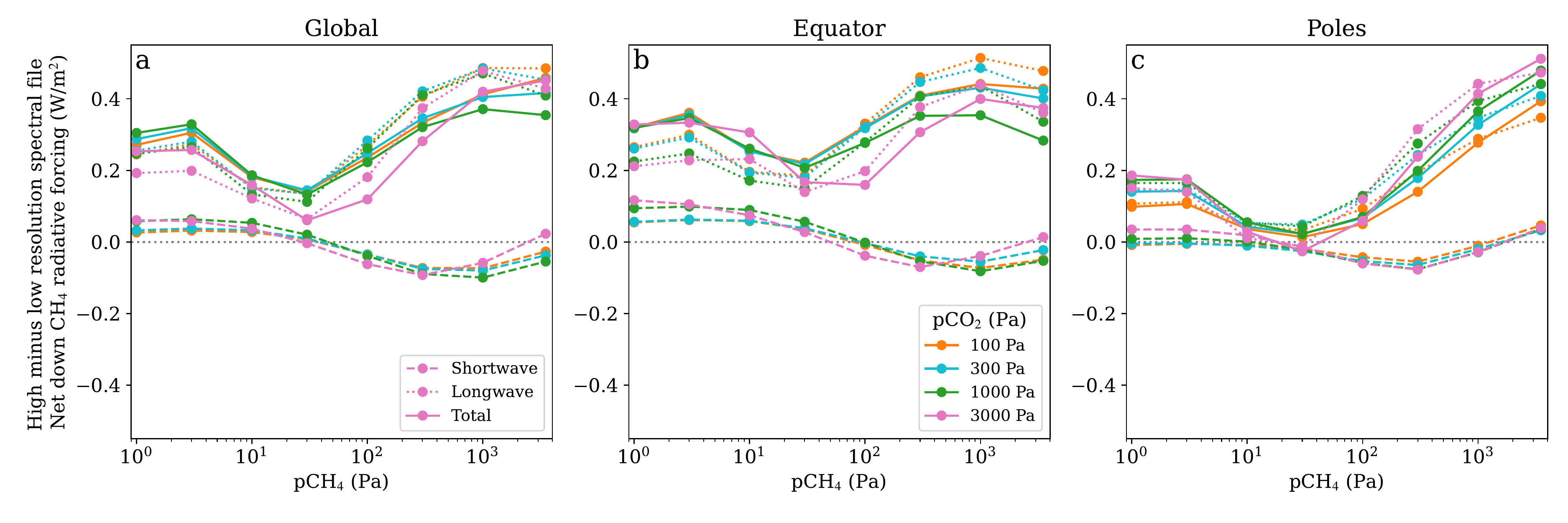}}

\caption{(a) shows the high resolution minus the low resolution spectral file global average net down radiative forcing at the tropopause for methane for shortwave (dashed), longwave (dotted) and their sum (solid). (b) and (c) show the same as (a) averaged over 10$\degree$S to 10$\degree$N as the equator in (b), and poleward of 70$\degree$N/S for the polar regions in (c). The high resolution spectral files for thermal and solar radiation are \texttt{sp\_lw\_350\_etw\_arcc10bar} and \texttt{sp\_sw\_280\_etw\_arcc10bar\_sun\_2.9gya} respectively, while the low resolution spectral files are \texttt{sp\_lw\_17\_etw\_arcc10bar} and \texttt{sp\_sw\_43\_etw\_arcc10bar\_sun\_2.9gya} for thermal and solar radiation respectively.
\label{fig:spec_file_test}
}
\end{figure}

Here, we make comparisons between spectral resolutions used in this study and a higher resolution spectral file treatment with 280 solar radiation bands \linebreak
(\texttt{sp\_sw\_280\_etw\_arcc10bar\_sun\_2.9gya}) and 350 thermal radiation bands \linebreak(\texttt{sp\_lw\_350\_etw\_arcc10bar}) to assess the accuracy of our radiative transfer calculation. The results of this are shown in \figurename~\ref{fig:spec_file_test}. Errors in both the shortwave and longwave forcing remain small compared to the trends observed in \figurename~\ref{fig:CH4_rad_forcings}, and thus will not affect the overall results.

\section{Calculations of heat fluxes in polar and equatorial regions}
\label{sec:lambert11}

Here, we outline the method used by \citeA{lambert11} to understand the net heat transport in and out of regions of the atmosphere. This method was initially designed for use in understanding heat transport by the atmosphere between regions of the atmosphere covering land and ocean. Here, we apply this to heat transport out of equatorial regions and into polar regions, with the heat flux out of the equatorial region, $\frac{\Delta A_E}{f_E}$, given as:

\begin{equation}
\label{eqn:equator}
    -\frac{\Delta A_E}{f_E} = \Delta N_E - \Delta U_E - \Delta U_{EA},
\end{equation}

where $\Delta A_E$ is the heat transport anomaly between the equatorial region and the rest of the planet. $f_E$ is the equatorial fraction of the global surface (spanning latitudes of 10\degree\,S to 10\degree\,N, $f_E=$17.4\%). $\Delta N_E$ is the net incoming radiation at the top-of-atmosphere, while $\Delta U_E$ is the net upward flux at the surface including both turbulent and radiative fluxes. $\Delta U_{EA}$ is the rate of heat storage by the atmosphere over the equatorial region, which is approximated by the globally averaged heat uptake by the atmosphere, $\Delta U_{GA}$:

\begin{equation}
    \Delta U_{GA} = \Delta N_G - \Delta U_G,
\end{equation}

where the subscript $G$ refers to the global averaged quantities discussed above. 

Equivalent expressions can be written for heat flux into the rest of the planet from the equatorial region, as well as the heat flux into the poles and the heat flux from the rest of the atmosphere into the poles:

\begin{equation}
\label{eqn:not_equator}
    \frac{\Delta A_E}{1-f_E} = \Delta N_{E'} - \Delta U_{E'} - \Delta U_{E'A},
\end{equation}

\begin{equation}
\label{eqn:pole}
    \frac{\Delta A_P}{f_P} = \Delta N_{P} - \Delta U_{P} - \Delta U_{PA},
\end{equation}

\begin{equation}
\label{eqn:not_pole}
    \frac{\Delta A_P}{1-f_P} = \Delta N_{P'} - \Delta U_{P'} - \Delta U_{P'A},
\end{equation}

where subscript $P$ represents the polar regions, while $E'$ and $P'$ are the rest of the world excluding the equatorial and polar regions respectively.

\begin{figure}
\centering
 \makebox[\textwidth][c]{\noindent\includegraphics[width=0.9\textwidth]{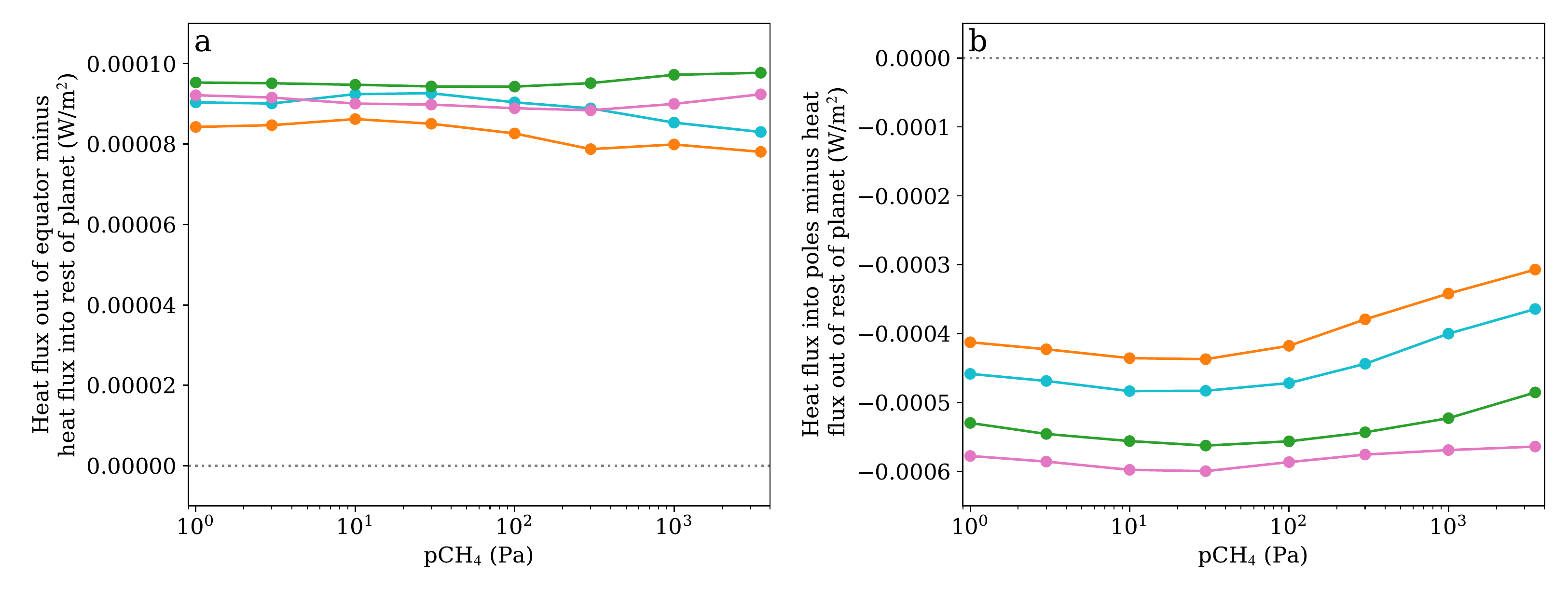}}

\caption{(a) shows the difference between the heat flux leaving the equatorial region and the heat flux into the rest of the atmosphere from the equator from Equation~\ref{eqn:equator} and Equation~\ref{eqn:not_equator}. Similarly (b) shows the difference between the heat flux entering the polar  regions and the heat flux from the rest of the atmosphere from Equation~\ref{eqn:pole} and Equation~\ref{eqn:not_pole}. The equatorial region is considered as spanning latitudes of 10\degree\,S to 10\degree\,N, while the polar regions are 70\degree\,S/N to the pole.
\label{fig:heat_flux_test}
}
\end{figure}

This method can be validated by calculating the difference of the heat flux between the equatorial/polar regions and the rest of the atmosphere. These are shown in \figurename~\ref{fig:heat_flux_test}, with the differences more than 1000 times smaller than 1 W/m$^2$, showing the self consistency of the method and that it is suitable for applying to understanding heat transport between regions of the atmosphere other than just those over land and sea.

\section{Methane radiative forcing: albedo dependence}
\label{sec:RF_albedo}

\begin{figure}
\centering
 \makebox[\textwidth][c]{\noindent\includegraphics[width=\textwidth]{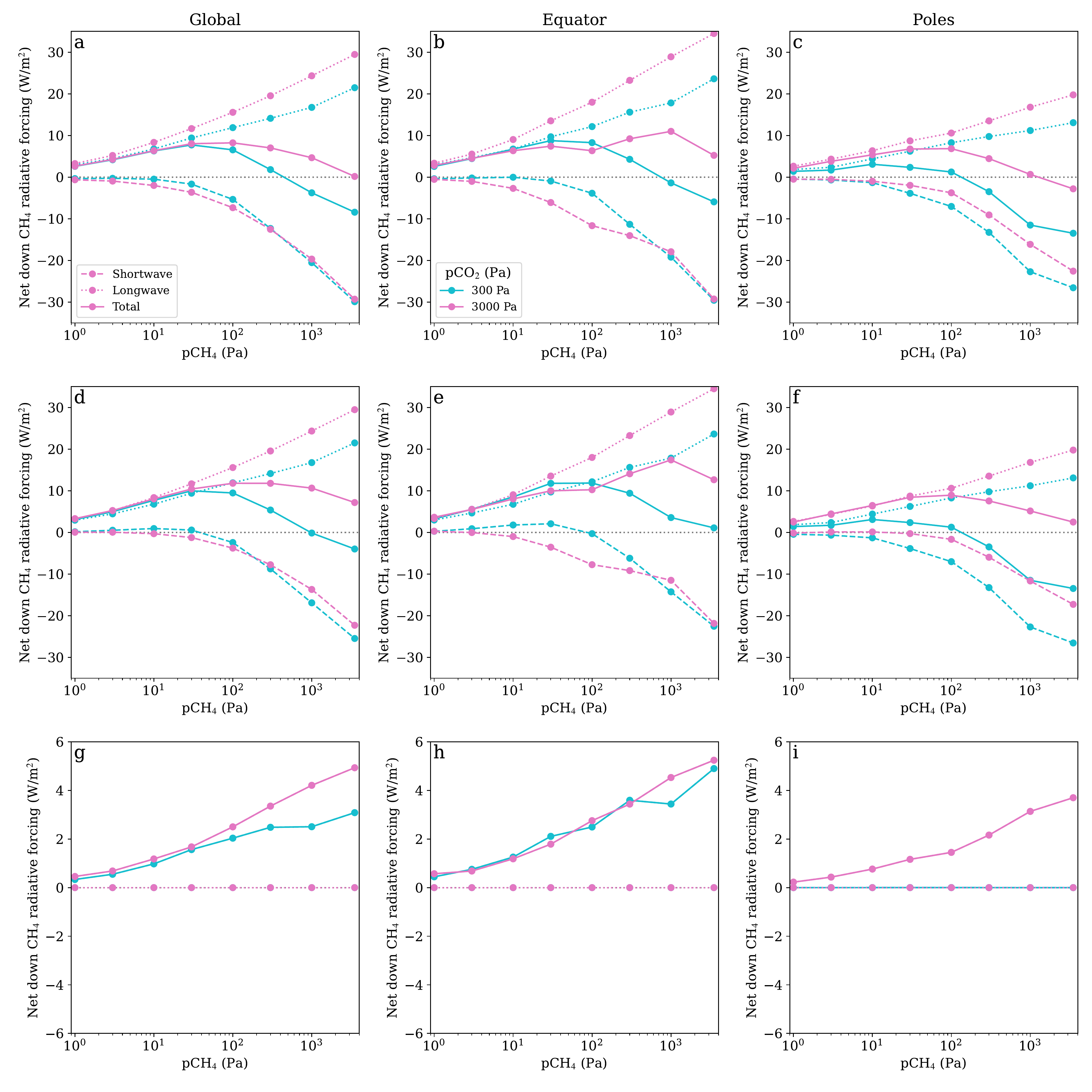}}

\caption{(a) shows the global average net down radiative forcing at the tropopause for methane for shortwave (dashed), longwave (dotted) and their sum (solid), for an ocean surface with an albedo of 0.07. (b) and (c) show the same as (a) averaged over 10$\degree$S to 10$\degree$N as the equator in (b), and poleward of 70$\degree$N/S for the polar regions in (c). These are calculated for just one model time step. (d), (e) and (f) shows the same as (a), (b) and (c) respectively, but for an ocean surface albedo of 0.533. (g), (h) and (i) show the difference between a), (b) and (c), and (d), (e) and (f), with a positive value indicating an increase in radiative forcing with the higher albedo of 0.533.
\label{fig:RF_albedo}
}
\end{figure}

To compare the effect of surface albedo on methane radiaitve forcing, \figurename~\ref{fig:RF_albedo} shows a sensitivity test for two surface albedos of 0.07 (case used in this study) and 0.533, which represents a high value for albedo produced by sand in the near infrared. \figurename~\ref{fig:RF_albedo} shows methane radiative forcing after running the model for one additional timestep from the configurations discussed in the main body of text, hence the slight differences between \figurename~\ref{fig:CH4_rad_forcings} and \figurename~\ref{fig:RF_albedo}a-c. As these are calculated with no evolution of the climate state, there is no change in the longwave radiative forcing between the two cases \figurename~\ref{fig:RF_albedo}g-i, and the change in shortwave methane radiative forcing is zero at the pole due to this region being ice and ice albedo remains unchanged.

Increasing ocean albedo causes the shortwave methane radiative forcing to decrease, however, general trends in radiative forcings discussed in Section~\ref{sec:results} remain the same. At high p\ch{CH4} global temperatures may increase, leading to a less pronounced decline in temperatures compared to those found in \figurename~\ref{fig:T_surf}.

\acknowledgments
We would like to thank the three reviewers for their very helpful reviews that improved the manuscript. We would also like to thank Colin Goldblatt for providing helpful comments on the manuscript. JE-N would like to thank the Hill Family Scholarship. The Hill Family Scholarship has been generously supported by University of Exeter alumnus, and president of the University's US Foundation Graham Hill (Economic \& Political Development, 1992) and other donors to the US Foundation. Material produced using Met Office Software. AN, NM and TL gratefully acknowledge funding from a Leverhulme Trust Research Project Grant [RPG-2020-82]. SD and TL would to thank the John Templeton Foundation Grant [62220]. JM and IB acknowledge the support of a Met Office Academic Partnership secondment. We acknowledge use of the Monsoon2 system, a collaborative facility supplied under the Joint Weather and Climate Research Programme, a strategic partnership between the Met Office and the Natural Environment Research Council. This research made use of the ISCA High Performance Computing Service at the University of Exeter. This work was partly supported by a Science and Technology Facilities Council Consolidated Grant [ST/R000395/1]. This work was also supported by a UKRI Future Leaders Fellowship [grant number MR/T040866/1]. For the purpose of open access, the author(s) has applied a Creative Commons Attribution (CC BY) licence to any Author Accepted Manuscript version arising. We would also like to note the energy intensive nature of super-computing. We
estimate the final production runs needed for this paper resulted in
roughly 4 tCO2e emitted into the atmosphere.

\bibliography{CH4_archean}

\end{document}


%
%


\title{Supporting Information for ``Insert Title"}
%
%

%
%



\authors{J. K. Eager\affil{1},
        N. J. Mayne\affil{1},
        T. M. Lenton\affil{2},
        A. E. Nicholson\affil{1},
        S. J. Daines\affil{2},
        D. E. Sergeev\affil{3},
        F. H. Lambert\affil{3},
        R. J. Ridgway\affil{1},
        J. Manners\affil{2,4},
        I. A. Boutle\affil{1,4},
        and 
        K. Kohary\affil{1}
        }

\affiliation{1}{Physics and Astronomy, College of Engineering, Mathematics and Physical Sciences, University of Exeter, Exeter, EX4 4QL, UK}
\affiliation{2}{Global Systems Institute, University of Exeter, Exeter, EX4 4QE, UK}
\affiliation{3}{Mathematics, College of Engineering, Mathematics and Physical Sciences, University of Exeter, Exeter, EX4 4QF, UK}
\affiliation{4}{Met Office, FitzRoy Road, Exeter, EX1 3PB, UK}

%
%

%

\begin{article}

%
%

\noindent\textbf{Contents of this file}
\begin{enumerate}
\item Text S1 to Sx
\item Figures S1 to Sx
\item Tables S1 to Sx
\end{enumerate}
\noindent\textbf{Additional Supporting Information (Files uploaded separately)}
\begin{enumerate}
\item Captions for Datasets S1 to Sx
\item Captions for large Tables S1 to Sx (if larger than 1 page, upload as separate excel file)
\item Captions for Movies S1 to Sx
\item Captions for Audio S1 to Sx
\end{enumerate}

\noindent\textbf{Introduction}


\noindent\textbf{Text S1.}
%


\noindent\textbf{Data Set S1.} 


\noindent\textbf{Movie S1.} 


\noindent\textbf{Audio S1.} 
\end{article}
\clearpage